\documentclass[a4paper,11pt]{article}
\usepackage{makeidx}
\usepackage{graphicx}
\usepackage[english,activeacute]{babel}
\usepackage[latin1]{inputenc}
\usepackage{amsmath}
\usepackage{amsfonts}
\usepackage{amssymb}
\usepackage{latexsym}
\usepackage{colortbl}
\usepackage{multicol}
\usepackage{hyperref}
\usepackage{xcolor}
\usepackage{orcidlink}
\usepackage{mathrsfs}
\usepackage{algorithmic}
\usepackage{algorithm}

\newcommand{\N}{\mathbb{N}}

\usepackage{mathtools}

\DeclarePairedDelimiter\floor{\lfloor}{\rfloor}

\newtheorem{theorem}{Theorem}

\newtheorem{corollary}{Corollary}

\newtheorem{definition}{Definition}

\newtheorem{lemma}{Lemma}

\newtheorem{proposition}{Proposition}

\newenvironment{proof}[1][Proof]{\noindent\textbf{#1.} }{\ \rule{0.5em}{0.5em}}
\newcommand{\dst}{\displaystyle}
\begin{document}

\title{A dual watermaking scheme based on Sobolev type orthogonal moments for document authentication}
\author{{Alicia María Centurión-Fajardo %
\orcidlink{0000-0003-1572-2631}}$^{1}$, {Alberto Lastra \orcidlink{0000-0002-4012-6471}}$^{2}$ {Anier Soria-Lorente \orcidlink{0000-0003-3488-3094}}$^{3}$, \\
%EndAName
\\
$^{1}$Departamento de Contabilidad y Finanzas, Universidad de Granma\\
Prolongación de General Garc\'{i}a\\
85100 - Bayamo, Cuba\\
amcenturionf@gmail.com\\
$^{2}$Universidad de Alcalá. Departamento de Física y Matemáticas, \\
Ctra. Madrid-Barcelona, Km. 33,600\\
28805 - Alcalá de Henares, Madrid, Spain\\
alberto.lastra@uah.es\\
$^{3}$Departamento de Tecnología, Universidad de Granma\\
Ctra. de Bayamo-Manzanillo, Km. 17,500\\
85100 - Bayamo, Cuba\\
asorial@udg.co.cu, asorial1983@gmail.com
}

\maketitle

\begin{abstract}
A dual watermarking scheme based on Sobolev type orthogonal moments, Charlier and Meixner, is proposed based on different discrete measures. The existing relation through the connection formulas allows to provide with structure and recurrence relations, together with two difference equations satisfied by such families. Weighted polynomials derived from them are being applied in an embedding and extraction watermarking algorithm, comparing the results obtained in imperceptibly and robustness tests with other families of polynomials.

\vspace{0.3cm}

\textit{Key words and phrases.} Orthogonal polynomials, discrete Sobolev
polynomials, difference equation, watermarking 

\textrm{2010 AMS Subject Classification. 33C45, 33C47.}

{\footnotesize Corresponding author: Alberto Lastra}
\end{abstract}

%%%%%%%%%%%%%%%%%%%%%%%%%%%%%%%%%%%%%%%%%%%%%%%%%%%%%%%%%%%%%%%%%%%%%%%%%%%%%%%%%%%%%%%%%%%%%%%%%%%%%%%

%%%%%%%%%%%%%%%%%%%%%%%%%%%%%%%%%%%%%%%%%%%%%%%%%%%%%%%%%%%%%%%%%%%%%%%%%%%%%%%%%%%%%%%%%%%%%%%%%%%%%%%

\section{Introduction}

The present work is a continuation of~\cite{HLS22} in the study of the behavior of Sobolev-type orthogonal polynomials when applied to watermarking schemes. In that previous work, the sequence of orthogonal polynomials considered was of Krawtchouk-Sobolev type, whereas this work is devoted to show the results on a watermarking process for Charlier (resp. Meixner) Sobolev-type polynomials. More precisely, let $\alpha$ be a negative real number, $\lambda>0$ and $j\in\N$. We consider families of orthogonal polynomials with respect to the measure
$$\langle f(x) , g(x)  \rangle_{\lambda}=\sum_{x\ge 0}f(x)g(x)\frac{e^{-\mu}\mu^x}{\Gamma(x+1)}+\lambda\Delta^jf(\alpha)\Delta^jg(\alpha),$$
for some fixed $\mu>0$, and where $\Gamma(\cdot)$ stands for Gamma function (resp.
$$\langle f(x) , g(x)  \rangle_{\lambda}=\sum_{x\ge 0}f(x)g(x)\frac{\mu^x\Gamma(\gamma+x)}{\Gamma(\gamma)\Gamma(x+1)}+\lambda\Delta^jf(\alpha)\Delta^jg(\alpha),$$
for some fixed $\gamma>0$ and $0<\mu<1$). Here, $\Delta$ stands for the forward difference operator defined by $\Delta f(x)=f(x+1)-f(x)$.

In the case of Charlier polynomials, the measure is the Poisson distribution of probability, with $\alpha$ being chosen in such a way that the interval $(\alpha,\alpha+1)$ does not contain points of increase of the distribution. Other recent contributions on Sobolev-type Charlier polynomials are~\cite{HS2019}, where the authors deal with the particular case $j=1$ and providing a hypergeometric representation, ladder operators and two different versions of the linear difference equation of second order associated to these polynomials. In the recent work~\cite{domo}, the asymptotic analysis of a family of polynomials of this type is treated in detail.

Regarding the study of Sobolev-type Meixner polynomials involving discrete orthogonality measures, we refer to~\cite{agm, agmmb, khol, mb, mbtpmp}, and the references therein.

The study of inner products involving difference operators was initiated by H. Bavinck in~\cite{B1995,B1995gen,B1996}, as a discrete version of the Sobolev inner products constructed in terms of derivative operators, of non-standard nature. Other Sobolev-type discrete orthogonal polynomials have been treated in~\cite{R22} (see also the references therein) when studying Sobolev operators based on a general divided-difference operator $\mathbb{D}$, generalizing $q-$differentiation. In~\cite{gamamo}, a Sobolev-type inner product under the action of Hahn difference operator is studied, constructing ladder operators and also deducing a second order differential-difference equation satisfied by them.

We also cite~\cite{almarero, mapepi, me} as references in this direction, together with the recent survey on the topic~\cite{maxu}.

The main purpose of the present article is to use the two Sobolev-type measures in the procedure of watermarking an image, say $\mathcal{C}$, and compare the results obtained with other previous available tools constructed in terms of other families of orthogonal polynomials by means of imperceptibility tests (PSNR - Peak Signal to Noise Ratio) and robustness test (BET - Bit error rate). The procedure of watermarking the cover image $\mathcal{C}$ consists, roughly speaking, on embedding information in $\mathcal{C}$ obtaining a modified image $\mathcal{W}$, known as the watermarked image. $\mathcal{C}$ and $\mathcal{W}$ should remain close enough for an external viewer whereas the initial image has been modified. The procedure is based on the use of Charlier and Meixner Sobolev-type orthogonal polynomials, studied in the first part of the work. The algorithm to transform $\mathcal{C}$ into $\mathcal{W}$ is described in Section~\ref{seck}.

The results obtained in the watermarking process allow us to conclude that the watermarking process based on Charlier Sobolev and Meixner Sobolev polynomials possess a better answer to external attacks, showing a robustness level higher that the one obtained with other schemes considered in the literature.

The work is structured as follows: Section~\ref{secsob} recalls the main facts about Charlier and Meixner families of orthogonal polynomials, to be applied in the construction of certain Sobolev-type discrete orthogonal polynomials by means of the connection formulas provided in Section~\ref{S3-ConnForm}. A description of the recurrence relation and difference equations associated to such polynomials are stated in Section~\ref{secpral}. The definition of weighted Sobolev type polynomials in Section~\ref{sec3} and their properties justify their choice for application in watermarking processes in the second part of the paper. The procedure of watermarking an image via the embedding and extraction algorithms is described in a section devoted to results and discussion of the experimental results of the proposed scheme: imperceptibily test, robustness and image temper detection, when compared with others.

%%%%%%%%%%%%%%%%%%%%%%%%%%%%%%%%%%%%%%%%%%%%%%%%%%%%%%%%%%%%%%%%%%%%%%%%%%%%%%%%%%%%%%%%%%%%%%%%%%%%%%%

\section{Introductory results on Charlier and Meixner polynomials}\label{secsob}

In this section, we briefly describe the main definitions and results on Charlier and Meixner families of orthogonal polynomials. We have decided to present the results in a more compact way, valid for both families when particularizing the two elements in Table~\ref{tableRsEM}, for the sake of readability. We also recall some of the main properties of backward and forward difference operators ($\Delta$ and $\nabla$ respectively).

Let us fix one of the elements in the first row of Table~\ref{tableRsEM}, either Charlier polynomials $C_{n}^{(\mu)}$ or Meixner polynomials $M_n^{(\gamma,\mu)}$. See~\cite{Ismail05,KLS2010,nikiforov1991classical} for further details on the topic.

\begin{table}[H]
	\caption{Monic Charlier $C_n^{(\mu)}(x)$ and Meixner polynomials $M_n^{(\gamma,\mu)}(x)$}
	\centering
	\renewcommand{\arraystretch}{2.0}
	\begin{tabular}{|c|c|c|} % {|l|c|c|} 
		\hline
		$P_{n}(x)$ & \quad $C_n^{(\mu)}(x)$, $\mu>0$ & $M_n^{(\gamma,\mu)}(x)$, $\gamma>0$, $0<\mu<1$\\ \hline\hline
		$\rho(x)$ & $\dst\frac{e^{-\mu}\mu^x}{\Gamma(x+1)}$ & $\dst\frac{\mu^x\Gamma(\gamma + x)}{\Gamma(\gamma)\Gamma(x+1)}$\\ \hline
		$||P_{n}(x)||^2$ & $n!\mu^n$ & $\dst\frac{n!(\gamma)_n\mu^n}{(1-\mu)^{\gamma+2n}}$\\ \hline
		$\alpha_n$ & $n+\mu$ & $\dst\frac{n(1+\mu)+\mu\gamma}{1-\mu}$\\\hline
		$\beta_n$ & $n\mu$ & $\dst\frac{n\mu(n-1+\gamma)}{(\mu-1)^2}$
		\\ \hline
		$\sigma(x)$ & $x$ & $x$\\ \hline
		$\tau(x)$ & $\mu-x$ & $(\mu-1)x+\mu\gamma$ \\\hline
		$\lambda_n$ & $n$ & $(1-\mu)n$\\\hline
		$\widetilde{\alpha}_n$ & $0$ & $n\mu$
		\\ \hline
		$\widetilde{\beta}_n$ & $n\mu$ & $\dst\frac{n\mu(n-1+\gamma)}{1-\mu}$
		\\ \hline
	\end{tabular}
	\label{tableRsEM}
\end{table}

\begin{proposition}
	Let $\{P_{n}\}_{n\geq 0}$ be the sequence of monic classical discrete polynomials of degree $n$ of the first row at Table~\ref{tableRsEM}. The following statements hold.
	
	\begin{enumerate}
		\item The sequence $\{P_{n}\}_{n\geq 0}$ consists of monic polynomials orthogonal with respect to the inner product defined on the space of polynomials $\mathbb{P}$, 
		\begin{equation*}
		\left\langle f,g\right\rangle =\sum_{x\geq 0}f(x)g(x)\rho(x),
		\end{equation*}
		where $\rho(x)$ is defined by the corresponding column of the second row of Table \ref{tableRsEM}.

		\item Squared norm. For every $n\in \mathbb{N}$,  $||P_n||^2=\left\langle P_n,P_n\right\rangle$ is shown at Table \ref{tableRsEM}.
		
		\item The three term recurrence relation 
		\begin{equation}
		xP_{n}\left( x\right) =P_{n+1}\left( x\right) +\alpha
		_{n}P_{n}\left( x\right) +\beta _{n}P_{n-1}\left(
		x\right) ,\quad n\geq 0,  \label{ReR}
		\end{equation}%
		holds, for $\alpha_{n}$ and $\beta_{n}$ in the corresponding column of Table \ref{tableRsEM}.
		
		\item Structure relation. For every $n\in \mathbb{N}$,
		\begin{equation}
		[\sigma(x)+\tau(x)]\Delta P_n\left( x\right) =\widetilde{\alpha}_nP_{n}\left( x\right)
		+\widetilde{\beta}_{n}P_{n-1}\left(x\right),  \label{StruR1}
		\end{equation}
		where $\widetilde{\alpha}_n$ and $\widetilde{\beta}_{n}$ are determined by the corresponding column of Table \ref{tableRsEM}. Here, $\nabla$ stands for the backward difference
		operator defined by
		\begin{equation*}
			\nabla f(x)=f(x)-f(x-1),
		\end{equation*}
		and recursively
		\begin{equation*}
		\nabla^{n}f(x)=\nabla[\nabla^{n-1}f(x)],\quad n\in \mathbb{N}.
		\end{equation*}
		
		\item Orthogonality relation. Given $a<0$ 
		\begin{equation*}
			\left\langle {P_n,P_n} \right\rangle=\left\Vert
			P_{n}\right\Vert ^{2}\delta _{m,n},
		\end{equation*}%
		where by $\delta _{i,j}$ we denote the Kronecker delta function.
		
		\item Second order difference equation (hypergeometric type equation) 
		\begin{equation*}
			\sigma(x)\Delta \nabla P_{n}(x) + \tau(x)\Delta P_{n}(x) +
		\lambda_nP_{n}(x) =0,
		\end{equation*}	
		where $\sigma(x)$, $\tau(x)$ and $\lambda_n$ are determined by the corresponding row of  Table \ref{tableRsEM}, and $\Delta $ denotes the forward difference
		operator, defined by $\Delta f\left( x\right) =f\left( x+1\right) -f\left(
		x\right) $. An explicit subindex $\Delta_x$ or $\Delta_y$ indicates the variable to which the operator is applied, when the operator is applied to functions of several variables. The forward difference operator is defined recursively by
		\begin{equation}\label{forWop}
		\Delta ^{n}f(x)=\Delta [\Delta ^{n-1}f(x)],\quad n\in \mathbb{N}.
		\end{equation}
and satisfies the product rule
		\begin{equation}\label{forwardOpProd}
			\Delta [f(x)g(x)]=f(x)\Delta g(x)+g(x+1)\Delta f(x).
		\end{equation}
			\end{enumerate}
\end{proposition}

Concerning Charlier and Meixner families of orthogonal polynomials, a Christoffel-Darboux formula is also available.

\begin{proposition}[Christoffel-Darboux formula]
	Let $\{P_{n}\}_{n\geq 0}$ be one of the two sequences of monic classical discrete polynomials of degree $n$ determined in Table~\ref{tableRsEM}. Let $K_n$ denote the $n$-th reproducing kernel, defined by 
	\begin{equation*}
	K_{n}(x,y)=\sum_{k=0}^{n}\frac{P_{k}(x)P_{k}(y)}{||P_{k}||^{2}}.
	\end{equation*}
	Then, for all $n\in \mathbb{N}$, it holds that 
	\begin{equation}
	K_{n}(x,y)=\frac{P_{n+1}(x) P_{n}(y) -P_{n+1}(y)P_{n}(x) }{\left(
		x-y\right) ||P_{n}||^{2}}.  \label{CDarb}
	\end{equation}
\end{proposition}
Let us also fix the following notation on the iterative application of the forward difference operator $K_{n}(x,y)$ with respect to each variable: for all $(i,j)\in\N^2$ we write
\begin{equation}
K_{n}^{(i,j)}(x,y):=\Delta_{x}^{i}\left( \Delta_{y}^{j}K
_{n}\left( x,y\right) \right) =\sum_{k=0}^{n}\frac{%
	\Delta^{i}P_{k}(x)\Delta^{j}P_{k}(y)}{\left\Vert
	P_{k}\right\Vert ^{2}}.  \label{Kij}
\end{equation}
The following result can be obtained of similar way to \cite[Proposition 3]%
{Costas-2022} under small modifications. We only sketch its proof.
\begin{proposition}\label{S1-LemmaKernel0j}
Let $\{P_{n}\}_{n\geq 0}$ be as above.The following statement holds for
	every $n\in \mathbb{N}$, 
	\begin{equation}
	K_{n-1}^{\left( 0,j\right) }\left( x,y\right) = {\mathcal{A}}_n^{(j)}(x,y)P_{n}\left( x\right)+{\mathcal{B}}_n^{(j)}(x,y)P_{n-1}\left( x\right),  \label{Kernel0j}
	\end{equation}
	with
	\begin{equation*}
	{\mathcal{A}}_n^{(j)}(x,y) = \frac{j!}{\left\Vert P_{n-1}\right\Vert
		^{2}[x-y]_{j+1}}\sum_{k=0}^j\frac{\Delta^{k}P_{n-1}\left(
		y\right) }{k!}[x-y]_{k},
	\end{equation*}
	and 
	\begin{equation*}
	{\mathcal{B}}_n^{(j)}(x,y) = -\frac{j!}{\left\Vert P_{n-1}\right\Vert
		^{2}[x-y]_{j+1}}\sum_{k=0}^j\frac{\Delta^{k}P_{n}\left(
		y\right) }{k!}[x-y]_{k},
	\end{equation*}
	where, $[\cdot]_k$ denotes the falling factorial of order $k$, defined by,
	\begin{equation*}
		[x]_k=\begin{cases}
			1, & k=0,\\\\
			\dst\prod_{j=0}^{k-1}(x-j), & k\geq 1.
		\end{cases}
	\end{equation*}
\end{proposition}
\begin{proof}
	In fact, applying to \eqref{CDarb} the $j$-th finite difference \eqref{forWop} with
	respect to $y$ we obtain
	\begin{equation}
	K_{n-1}^{(0,j)}(x,y)=\dfrac{1} {\| P_{n-1}\|^{2}}%
	\left[P_{n}(x) \Delta_{y}^{j}\left(\frac{P_{n-1}(y)}{x-y}\right) -P_{n-1}(x)\Delta_{y}^{j} \left(\frac{P_{n}(y)}{x-y}\right)\right].  \label{K0jI}
	\end{equation}
	Using an analogue of the Leibniz's rule 
	\begin{equation*}
	\Delta^{n}\left[ f(x) g(x)\right] =\sum_{k=0}^{n}\binom{n}{k}%
	\Delta^{k}[f(x)] \Delta^{n-k}\left[ g( x+ k) \right]%
	,  \label{LR}
	\end{equation*}
	and 
	\begin{equation*}
	\Delta_{y}^{n}\left(\frac{1}{x-y}\right)=\frac{n!}{[x-y]_{n+1}},
	\end{equation*}
	we deduce 
	\begin{eqnarray*}
		\Delta_{y}^{j}\left(\frac{P_{n-1}(y)}{x-y}\right)
		&=&\sum_{k=0}^{j}\binom{j}{k}\Delta^{k}\left[ P_{n-1}( y) \right] \Delta_{y}^{j-k} \left( \frac{1}{x-y- k}\right) \\
		&=&\sum_{k=0}^{j}\frac{\Delta^{k}P_{n-1}( y)}{k!}%
		\frac{j!}{[x-y- k]_{j+1-k}}.
	\end{eqnarray*}
	Since 
	\begin{equation*}
	\frac{[x]_{n}}{[x]_{k}}=[x- k]_{n-k}\quad \mbox{if}\quad
	n\geq k,
	\end{equation*}
	we deduce 
	\begin{equation*}
		[x-y- k]_{j+1-k}=\frac{[x-y]_{j+1}}{[x-y]_{k}}.
	\end{equation*}
	Thus, 
	\begin{equation*}
	\Delta_{y}^{j}\left(\frac{P_{n-1}(y)}{x-y}\right) =%
	\frac{j!}{[x-y]_{j+1}}\sum_{0\leq k\leq j}\frac{%
	\Delta^{k}P_{n-1}(y)}{k!}[x-y]_{k}.
	\end{equation*}
	Therefore, from the above and \eqref{K0jI} we get \eqref{Kernel0j}.
\end{proof}

The connection formulas of the Sobolev-type polynomials and the classical ones is based on the following results.

\begin{proposition}
	\label{S1-LemmaKerneli2} Let $\{P_{n}\}_{n\geq 0}$ be as before. The following statements hold, for 	all $n,j\in \mathbb{N}$,
	\begin{equation}
		K_{n-1}^{(1,j)}(x,y) ={\mathcal{C}}_{1,n}(x,y)P_{n}(x)+{\mathcal{D}}_{1,n}(x,y)P_{n-1}(x),  \label{kernel1j}
	\end{equation}
	and
	\begin{equation}
		K_{n-1}^{(2,j)}(x,y) ={\mathcal{C}}_{2,n}(x,y)P_{n}(x)+{\mathcal{D}}_{2,n}(x,y)P_{n-1}(x),  \label{kernel2j}
	\end{equation}
	with%
	\begin{equation*}
	{\mathcal{C}}_{1,n}(x,y)=\Delta{\mathcal{A}}_{n}^{(j)}(x,y)+\widetilde{\alpha}_n\,\Theta(x)\,{\mathcal{A}}_{n}^{(j)}(x+1,y)-\widetilde{\beta}_{n-1}\,\beta _{n-1}^{-1}\,\Theta(x)\,{\mathcal{B}}_{n}^{(j)}(x+1,y),
	\end{equation*}
	\begin{multline*}
	{\mathcal{D}}_{1,n}(x,y)=\widetilde{\beta}_{n-1}\,\Theta(x)\,(\beta _{n-1}^{-1}\,x-\alpha
		_{n-1}\beta _{n-1}^{-1}){\mathcal{B}}_{n}^{(j)}(x+1,y)+\widetilde{\beta}_n\,\Theta(x)\,{\mathcal{A}}_{n}^{(j)}(x+1,y)\\+\widetilde{\alpha}_{n-1}\,\Theta(x)\,{\mathcal{B}}_{n}^{(j)}(x+1,y)+\Delta{\mathcal{B}}_{n}^{(j)}(x,y).
	\end{multline*}
	and
	\begin{equation*}
	{\mathcal{C}}_{2,n}(x,y)=\Delta{\mathcal{C}}_{1,n}^{(j)}(x,y)+\widetilde{\alpha}_n\,\Theta(x)\,{\mathcal{C}}_{1,n}^{(j)}(x+1,y)-\widetilde{\beta}_{n-1}\,\beta _{n-1}^{-1}\,\Theta(x)\,{\mathcal{D}}_{1,n}^{(j)}(x+1,y),
	\end{equation*}
	\begin{multline*}
	{\mathcal{D}}_{2,n}(x,y)=\widetilde{\beta}_{n-1}\,\Theta(x)\,(\beta _{n-1}^{-1}\,x-\alpha
	_{n-1}\beta _{n-1}^{-1})\,{\mathcal{D}}_{1,n}^{(j)}(x+1,y)+\widetilde{\beta}_n\,\Theta(x)\,{\mathcal{C}}_{1,n}^{(j)}(x+1,y)\\+\widetilde{\alpha}_{n-1}\,\Theta(x)\,{\mathcal{D}}_{1,n}^{(j)}(x+1,y)+\Delta{\mathcal{D}}_{1,n}^{(j)}(x,y).
	\end{multline*}
	and where
	\begin{equation*}
		\Theta(x)=(\sigma(x)+\tau(x))^{-1}.
	\end{equation*}
\end{proposition}

\begin{proof}
	We first apply the forward operator $\Delta$ with respect to $x$ variable in the expression (\ref{Kernel0j}), together with the property (\ref{forwardOpProd}). This	yields%
	\begin{multline*}
	K_{n-1,q}^{(1,j)}(x,y)=P_n(x)\Delta{\mathcal{A}}_{n}^{(j)}(x,y)+{\mathcal{A}}_{n}^{(j)}(x+1,y)\Delta P_n(x)\\
	+P_{n-1}(x)\Delta{\mathcal{B}}_{n}^{(j)}(x,y)+{\mathcal{B}}_{n}^{(j)}(x+1,y)\Delta P_{n-1}(x).
	\end{multline*}
	The structure relation \eqref{StruR1} guarantees that
	\begin{multline*}
	K_{n-1,q}^{(1,j)}(x,y)=P_n(x)\Delta{\mathcal{A}}_{n}^{(j)}(x,y)+\frac{\widetilde{\alpha}_n\,{\mathcal{A}}_{n}^{(j)}(x+1,y)}{\sigma(x)+\tau(x)}\, P_n(x)+\frac{\widetilde{\beta}_n\,{\mathcal{A}}_{n}^{(j)}(x+1,y)}{\sigma(x)+\tau(x)}\, P_{n-1}(x)\\
	+P_{n-1}(x)\Delta{\mathcal{B}}_{n}^{(j)}(x,y)+\frac{\widetilde{\alpha}_{n-1}\,{\mathcal{B}}_{n}^{(j)}(x+1,y)}{\sigma(x)+\tau(x)}\, P_{n-1}(x)+\frac{\widetilde{\beta}_{n-1}\,{\mathcal{B}}_{n}^{(j)}(x+1,y)}{\sigma(x)+\tau(x)}\, P_{n-2}(x).
	\end{multline*}
A direct application of \eqref{ReR} concludes with statement \eqref{kernel1j}. An analogous argument can be followed to obtain (\ref{kernel2j}) from (\ref{kernel1j}).
\end{proof}

\section{Sobolev-type polynomials and connection formulas}

\label{S3-ConnForm}

%%%%%%%%%%%%%%%%%%%%%%%%%%%%%%%%%%%%%%%%%%%%%%%%%%%%%%%%%%%%%%%%%%%%%%%%%%%%%%%%%%%%%%%%%%%%%%%%%%%%%%%

%%%%%%%%%%%%%%%%%%%%%%%%%%%%%%%%%%%%%%%%%%%%%%%%%%%%%%%%%%%%%%%%%%%%%%%%%%%%%%%%%%%%%%%%%%%%%%%%%%%%%%%

In this section we introduce the Sobolev type polynomials of higher order $\{\mathbb{S}_{n}(x)\}_{n\geq 0}$, which are orthogonal with
respect to the Sobolev-type inner product%
\begin{equation}
\left\langle f,g\right\rangle _{\lambda}
=\sum_{x\geq 0}f(x)g(x)\rho(x)+\lambda\Delta^jf(\alpha)\,\Delta^jg(\alpha),  \label{piSob}
\end{equation}
where $\alpha\in\mathbb{R}_{-}$, $\lambda\in\mathbb{R}^{+}$ and $%
j\in\mathbb{N}$ are fixed. Here, we maintain the choice available from Table~\ref{tableRsEM}, allowing to construct both Sobolev-type polynomials from Charlier or from Meixner polynomials. Notation is also maintained here.

The connection between  the Sobolev type
polynomials of higher order $\{\mathbb{S}_{n}(x)\}_{n\geq 0}$ and the classical discrete polynomials $\{P_{n}\}_{n\geq 0}$ of degree $n$ is stated in the next result. The difference equations satisfied by such polynomials lean on this connection formula.

\begin{proposition}
	Let $\{\mathbb{S}_{n}(x)\}_{n\geq 0}$ be the sequence of Sobolev type polynomials of degree $n$. Then, the following statements
	hold for $n\geq 1$ 
	\begin{equation}
	\mathbb{S}_{n}(x)=P_{n}(x)-\lambda \frac{\Delta^jP_{n}(\alpha)}{1+\lambda K_{n-1}^{(j,j)}(\alpha ,\alpha )}%
	K_{n-1}^{(0,j)}(x,\alpha ).  \label{ConxF1}
	\end{equation}
\end{proposition}

\begin{proof}
The first part of Section 2 in \cite{MR1990} can be applied here to conclude \eqref{ConxF1}.
\end{proof}

\begin{corollary}
	\label{DqDq2HS} Let $\{\mathbb{S}_{n}(x)\}_{n\geq 0}$ be the sequence of Sobolev type polynomials of degree $n$. Then, the following
	statement holds for every $n\geq 1$ and any $\ell\ge0$.
%	\begin{equation*}
%	\Delta\mathbb{S}_{n}(x)=\Delta P_{n}(x)-\lambda \frac{\Delta^jP_{n}(\alpha)}{1+\lambda K_{n-1}^{(j,j)}(\alpha ,\alpha )}%
%	K_{n-1}^{(1,j)}(x,\alpha ),
%	\end{equation*}
		\begin{equation*}
	\Delta^{\ell}\mathbb{S}_{n}(x)=\Delta^{\ell} P_{n}(x)-\lambda \frac{\Delta^jP_{n}(\alpha)}{1+\lambda K_{n-1}^{(j,j)}(\alpha ,\alpha )}%
	K_{n-1}^{(\ell,j)}(x,\alpha ).
	\end{equation*}
\end{corollary}

\begin{lemma}
	Let $\{\mathbb{S}_{n}(x)\}_{n\geq 0}$ be the sequence of Sobolev type polynomials of degree $n$. For every $n\geq 1$, one has 
	\begin{equation}
	\mathbb{S}_{n}(x)={\mathcal{E}}_{1,n}(x)P_{n}(x)+{\mathcal{F}}_{1,n}(x)P_{n-1}(x),  \label{ConexF_I}
	\end{equation}%
	where 
	\begin{equation}
	{\mathcal{E}}_{1,n}(x)=1-\lambda \frac{\Delta^jP_{n}(\alpha)}{1+\lambda K_{n-1}^{(j,j)}(\alpha ,\alpha )}{\mathcal{A}}%
	_{n}^{(j)}(x,\alpha ),  \label{EEq}
	\end{equation}%
	and 
	\begin{equation}
	{\mathcal{F}}_{1,n}(x)=-\lambda \frac{\Delta^jP_{n}(\alpha)}{1+\lambda K_{n-1}^{(j,j)}(\alpha ,\alpha )}{\mathcal{B}}%
	_{n}^{(j)}(x,\alpha ),  \label{FEq}
	\end{equation}%
%	\begin{equation*}
%	\mathbb{S}_{n}(x)=P_{n}(x)-\lambda \frac{\Delta^jP_{n-j}(\alpha)}{1+\lambda K_{n-1}^{(j,j)}(\alpha ,\alpha )}%
%	K_{n-1}^{(0,j)}(x,\alpha ),\quad n\geq 0.
%	\end{equation*}
\end{lemma}

\begin{proof}
The result follows directly in view of (\ref{ConxF1}) and Proposition \ref{S1-LemmaKernel0j}.
\end{proof}

\begin{lemma}
	Let $\{\mathbb{S}_{n}(x)\}_{n\geq 0}$ be the sequence of Sobolev type polynomials of degree $n$. Then, the following statements
	hold for $n\geq 1$, 
	\begin{equation}
	\mathbb{S}_{n-1}(x)={\mathcal{E}}_{2,n}(x)P_{n}(x)+{\mathcal{F}}_{2,n}(x)P_{n-1}(x),  \label{ConexF_II}
	\end{equation}%
	where 
	\begin{equation*}
	{\mathcal{E}}_{2,n}(x)=-\frac{{\mathcal{F}}_{1,n-1}(x)}{\beta_{n-1}},
	\end{equation*}%
	and 
	\begin{equation*}
	{\mathcal{F}}_{2,n}(x)={\mathcal{E}}_{1,n-1}(x)-(x-\alpha_{n-1}){\mathcal{E}}_{2,n}(x),
	\end{equation*}
\end{lemma}

\begin{proof}
It is a direct consequence of the previous Lemma and the recurrence relation \eqref{ReR}.
\end{proof}

\begin{lemma}
	\label{detxi} Let $\{\mathbb{S}_{n}(x)\}_{n\geq 0}$ be the sequence of Sobolev type polynomials of degree $n$. Then, the following
	statements hold for $n\geq 1$%
	\begin{equation}
	\Xi _{1,n}(x)P_{n}(x)=
	\begin{vmatrix}
	\mathbb{S}_{n}(x) & \mathbb{S}_{n-1}(x) \\ 
	{\mathcal{F}}_{1,n}(x) & {\mathcal{F}}_{2,n}(x)%
	\end{vmatrix}%
	,  \label{ConexF_III}
	\end{equation}%
	and 
	\begin{equation}
	\Xi _{1,n}(x)P_{n-1}(x)=-%
	\begin{vmatrix}
	\mathbb{S}_{n}(x) & \mathbb{S}_{n-1}(x) \\ 
	{\mathcal{E}}_{1,n}(x) & {\mathcal{E}}_{2,n}(x)%
	\end{vmatrix}%
	,  \label{ConexF_IV}
	\end{equation}%
	where 
	\begin{equation*}
	\Xi _{1,n}(x)=%
	\begin{vmatrix}
	{\mathcal{E}}_{1,n}(x) & {\mathcal{E}}_{2,n}(x) \\ 
	{\mathcal{F}}_{1,n}(x) & {\mathcal{F}}_{2,n}(x)%
	\end{vmatrix}.
	\end{equation*}%
\end{lemma}

\begin{proof}
Multiply (\ref{ConexF_I}) by ${\mathcal{F}}_{2,n}(x)$ and %
	(\ref{ConexF_II}) by $-{\mathcal{F}}_{1,n}(x)$. The sum and simplification of the resulting formula allows to deduce (\ref{ConexF_III}). Statement (\ref{ConexF_IV}) can be achieved in an analogous manner.
\end{proof}

%%%%%%%%%%%%%%%%%%%%%%%%%%%%%%%%%%%%%%%%%%%%%%%%%%%%%%%%%%%%%%%%%%%%%%%%%%%%%%%%%%%%%%%%%%%%%%%%%%%%%%%

%%%%%%%%%%%%%%%%%%%%%%%%%%%%%%%%%%%%%%%%%%%%%%%%%%%%%%%%%%%%%%%%%%%%%%%%%%%%%%%%%%%%%%%%%%%%%%%%%%%%%%%

\begin{lemma}
	\label{DqHS} Let $\{\mathbb{S}_{n}(x)\}_{n\geq 0}$ be the sequence of Sobolev type polynomials of degree $n$. Then, the following
	statement holds for $n\geq 1$, 
	\begin{equation}
	\Delta\mathbb{S}_{n}(x)={\mathcal{E}}_{3,n}(x)P_{n}(x)+{\mathcal{F}}_{3,n}(x)P_{n-1}(x),  \label{DqHsE3F3}
	\end{equation}%
	where 
	\begin{equation*}
	{\mathcal{E}}_{3,n}(x)=\widetilde{\alpha}_n\,\Theta(x)-\lambda \frac{\Delta^jP_{n}(\alpha)}{1+\lambda K_{n-1}^{(j,j)}(\alpha ,\alpha )}{\mathcal{C}}_{1,n}(x,\alpha
	),
	\end{equation*}%
	and 
	\begin{equation*}
	{\mathcal{F}}_{3,n}(x)=\widetilde{\beta}_n\,\Theta(x)-\lambda \frac{\Delta^jP_{n}(\alpha)}{1+\lambda K_{n-1}^{(j,j)}(\alpha ,\alpha )}{\mathcal{D}}_{1,n}(x,\alpha ).
	\end{equation*}%
\end{lemma}

%%%%%%%%%%%%%%%%%%%%%%%%%%%%%%%%%%%%%%%%%%%%%%%%%%%%%%%%%%%%%%%%%%%%%%%%%%%%%%%%%%%%%%%%%%%%%%%%%%%%%%%

%%%%%%%%%%%%%%%%%%%%%%%%%%%%%%%%%%%%%%%%%%%%%%%%%%%%%%%%%%%%%%%%%%%%%%%%%%%%%%%%%%%%%%%%%%%%%%%%%%%%%%%

\begin{proof}
	From Corollary \ref{DqDq2HS} applied with $\ell=1$, the structure relation \eqref{StruR1} and Proposition \ref{S1-LemmaKerneli2}, we conclude the result.
\end{proof}
%\begin{equation}
%[\sigma(x)+\tau(x)]\Delta P_n\left( x\right) =\widetilde{\alpha}_nP_{n}\left( x\right)
%+\widetilde{\beta}_{n}P_{n-1}\left(x\right),  \label{StruR1}
%\end{equation}

%%%%%%%%%%%%%%%%%%%%%%%%%%%%%%%%%%%%%%%%%%%%%%%%%%%%%%%%%%%%%%%%%%%%%%%%%%%%%%%%%%%%%%%%%%%%%%%%%%%%%%%

%%%%%%%%%%%%%%%%%%%%%%%%%%%%%%%%%%%%%%%%%%%%%%%%%%%%%%%%%%%%%%%%%%%%%%%%%%%%%%%%%%%%%%%%%%%%%%%%%%%%%%%

\begin{proposition}
	\label{STHST} The Sobolev type polynomials $\mathbb{S}_{n}(x) $ of degree $n$ satisfy the following structure relation for all $%
	n\geq 1$,%
	\begin{equation*}
	\Xi _{1,n}(x)\Delta\mathbb{S}_{n}(x)={\mathcal{E}}_{4,n}(x)\mathbb{S}_{n}(x)+{\mathcal{F}}_{4,n}(x)\mathbb{S}_{n-1}(x),
	\end{equation*}%
	where 
	\begin{equation*}
	{\mathcal{E}}_{4,n}(x)=-%
	\begin{vmatrix}
	{\mathcal{E}}_{2,n}(x) & {\mathcal{E}}_{3,n}(x) \\ 
	{\mathcal{F}}_{2,n}(x) & {\mathcal{F}}_{3,n}(x)%
	\end{vmatrix}%
	,
	\end{equation*}%
	and 
	\begin{equation*}
	{\mathcal{F}}_{4,n}(x)=%
	\begin{vmatrix}
	{\mathcal{E}}_{1,n}(x) & {\mathcal{E}}_{3,n}(x) \\ 
	{\mathcal{F}}_{1,n}(x) & {\mathcal{F}}_{3,n}(x)%
	\end{vmatrix}.
	\end{equation*}
\end{proposition}

%%%%%%%%%%%%%%%%%%%%%%%%%%%%%%%%%%%%%%%%%%%%%%%%%%%%%%%%%%%%%%%%%%%%%%%%%%%%%%%%%%%%%%%%%%%%%%%%%%%%%%%

%%%%%%%%%%%%%%%%%%%%%%%%%%%%%%%%%%%%%%%%%%%%%%%%%%%%%%%%%%%%%%%%%%%%%%%%%%%%%%%%%%%%%%%%%%%%%%%%%%%%%%%

\begin{proof}
	The application of Lemma \ref{detxi} and Lemma \ref{DqHS}, guarantees that%
	\begin{multline*}
	\begin{vmatrix}
	\mathbb{S}_{n}(x) & \mathbb{S}_{n-1}(x) \\ 
	{\mathcal{F}}_{1,n}(x) & {\mathcal{F}}_{2,n}(x)%
	\end{vmatrix}%
	{\mathcal{E}}_{3,n}(x)-%
	\begin{vmatrix}
	\mathbb{S}_{n}(x) & \mathbb{S}_{n-1}(x) \\ 
	{\mathcal{E}}_{1,n}(x) & {\mathcal{E}}_{2,n}(x)%
	\end{vmatrix}%
	{\mathcal{F}}_{3,n}(x)\\={\mathcal{E}}_{3,n}(x){\mathcal{F}}_{2,n}(x)\mathbb{S}_{n}(x)-{\mathcal{%
			E}}_{3,n}(x){\mathcal{F}}_{1,n}(x)\mathbb{S}_{n-1}(x) \\
	\qquad \qquad \qquad \qquad \qquad \qquad -{\mathcal{E}}_{2,n}(x){\mathcal{%
			F}}_{3,n}(x)\mathbb{S}_{n}(x)+{\mathcal{E}}_{1,n}(x){\mathcal{F}}_{3,n}(x)%
	\mathbb{S}_{n-1}(x)\\=-%
	\begin{vmatrix}
	{\mathcal{E}}_{2,n}(x) & {\mathcal{E}}_{3,n}(x) \\ 
	{\mathcal{F}}_{2,n}(x) & {\mathcal{F}}_{3,n}(x)%
	\end{vmatrix}%
	\mathbb{S}_{n}(x)+%
	\begin{vmatrix}
	{\mathcal{E}}_{1,n}(x) & {\mathcal{E}}_{3,n}(x) \\ 
	{\mathcal{F}}_{1,n}(x) & {\mathcal{F}}_{3,n}(x)%
	\end{vmatrix}%
	\mathbb{S}_{n-1}(x).
	\end{multline*}%
%	\begin{eqnarray*}
%		&&={\mathcal{E}}_{3,n}(x){\mathcal{F}}_{2,n}(x)\mathbb{S}_{n}(x)-{\mathcal{%
%				E}}_{3,n}(x){\mathcal{F}}_{1,n}(x)\mathbb{S}_{n-1}(x) \\
%		&&\qquad \qquad \qquad \qquad \qquad \qquad -{\mathcal{E}}_{2,n}(x){\mathcal{%
%				F}}_{3,n}(x)\mathbb{S}_{n}(x)+{\mathcal{E}}_{1,n}(x){\mathcal{F}}_{3,n}(x)%
%		\mathbb{S}_{n-1}(x)
%	\end{eqnarray*}%
%	\begin{equation*}
%	
%	\end{equation*}
	This concludes the result.
\end{proof}

%%%%%%%%%%%%%%%%%%%%%%%%%%%%%%%%%%%%%%%%%%%%%%%%%%%%%%%%%%%%%%%%%%%%%%%%%%%%%%%%%%%%%%%%%%%%%%%%%%%%%%%

%%%%%%%%%%%%%%%%%%%%%%%%%%%%%%%%%%%%%%%%%%%%%%%%%%%%%%%%%%%%%%%%%%%%%%%%%%%%%%%%%%%%%%%%%%%%%%%%%%%%%%%

\begin{lemma}
	\label{Dq2HS} Let $\{\mathbb{S}_{n}(x)\}_{n\geq 0}$ be the sequence of Sobolev type polynomials of degree $n$. Then, the following
	statements hold for all $n\geq 1$%
	\begin{equation}
	\Delta^{2}\mathbb{S}_{n}(x)={\mathcal{E}}_{5,n}(x)P_{n}(x)+{\mathcal{F}}_{5,n}(x)P_{n-1}(x),  \label{Dq2Hs}
	\end{equation}%
	where 
	\begin{equation*}
	{\mathcal{E}}_{5,n}(x)=\Theta_{1,n}(x)-\lambda \frac{\Delta^jP_{n}(\alpha)}{1+\lambda K_{n-1}^{(j,j)}(\alpha ,\alpha )}{\mathcal{C}}_{2,n}(x,\alpha ),
	\end{equation*}%
	and%
	\begin{equation*}
	{\mathcal{F}}_{5,n}(x)=\Theta_{2,n}(x)-\lambda \frac{\Delta^jP_{n}(\alpha)}{1+\lambda K_{n-1}^{(j,j)}(\alpha ,\alpha )}{\mathcal{D}}_{2,n}(x,\alpha ),
	\end{equation*}
	with,
	\begin{equation*}
	\Theta_{1,n}(x)=\widetilde{\alpha}_n\,\Delta \Theta(x)+\widetilde{\alpha}_n^2\,\Theta(x)\,\Theta(x+1)-\widetilde{\beta}_{n-1}\beta _{n-1}^{-1}\,\widetilde{\beta}_n\,\Theta(x+1),
	\end{equation*}
	and
	\begin{multline*}
	\Theta_{2,n}(x)=\widetilde{\beta}_n\,\Delta \Theta(x)+\widetilde{\alpha}_n\widetilde{\beta}_n\,\Theta(x)\,\Theta(x+1)+\widetilde{\alpha}_{n-1}\,\widetilde{\beta}_n\,\Theta(x+1)\\+\widetilde{\beta}_{n-1}\,\beta _{n-1}^{-1}\,\widetilde{\beta}_n\,\Theta(x+1)\,(x-\alpha
	_{n-1}).
	\end{multline*}
\end{lemma}

%%%%%%%%%%%%%%%%%%%%%%%%%%%%%%%%%%%%%%%%%%%%%%%%%%%%%%%%%%%%%%%%%%%%%%%%%%%%%%%%%%%%%%%%%%%%%%%%%%%%%%%

%%%%%%%%%%%%%%%%%%%%%%%%%%%%%%%%%%%%%%%%%%%%%%%%%%%%%%%%%%%%%%%%%%%%%%%%%%%%%%%%%%%%%%%%%%%%%%%%%%%%%%%

\begin{proof}
	It is straightforward to check the result from Corollary \ref%
	{DqDq2HS} for $\ell=2$, the structure relation (\ref{StruR1}), the product formula in (\ref{forwardOpProd}) and Proposition \ref{S1-LemmaKerneli2}.
\end{proof}

%%%%%%%%%%%%%%%%%%%%%%%%%%%%%%%%%%%%%%%%%%%%%%%%%%%%%%%%%%%%%%%%%%%%%%%%%%%%%%%%%%%%%%%%%%%%%%%%%%%%%%%

%%%%%%%%%%%%%%%%%%%%%%%%%%%%%%%%%%%%%%%%%%%%%%%%%%%%%%%%%%%%%%%%%%%%%%%%%%%%%%%%%%%%%%%%%%%%%%%%%%%%%%%

\begin{proposition}
	\label{2SR} The Sobolev type polynomials $\{\mathbb{S}_{n}(x)\}_{n\geq 0}$	of degree $n$ satisfy the following relation for all $n\geq 1$, 
	\begin{equation*}
	\Xi _{1,n}(x)\Delta^{2}\mathbb{S}_{n}(x)={\mathcal{E}}_{6,n}(x)%
	\mathbb{S}_{n}(x)+{\mathcal{F}}_{6,n}(x)\mathbb{S}_{n-1}(x),
	\end{equation*}%
	where 
	\begin{equation*}
	{\mathcal{E}}_{6,n}(x)=-%
	\begin{vmatrix}
	{\mathcal{E}}_{2,n}(x) & {\mathcal{E}}_{5,n}(x) \\ 
	{\mathcal{F}}_{2,n}(x) & {\mathcal{F}}_{5,n}(x)%
	\end{vmatrix}%
	,
	\end{equation*}%
	and 
	\begin{equation*}
	{\mathcal{F}}_{6,n}(x)=%
	\begin{vmatrix}
	{\mathcal{E}}_{1,n}(x) & {\mathcal{E}}_{5,n}(x) \\ 
	{\mathcal{F}}_{1,n}(x) & {\mathcal{F}}_{5,n}(x)%
	\end{vmatrix}%
	.
	\end{equation*}
\end{proposition}

%%%%%%%%%%%%%%%%%%%%%%%%%%%%%%%%%%%%%%%%%%%%%%%%%%%%%%%%%%%%%%%%%%%%%%%%%%%%%%%%%%%%%%%%%%%%%%%%%%%%%%%

%%%%%%%%%%%%%%%%%%%%%%%%%%%%%%%%%%%%%%%%%%%%%%%%%%%%%%%%%%%%%%%%%%%%%%%%%%%%%%%%%%%%%%%%%%%%%%%%%%%%%%%

\begin{proof}
It is a direct consequence of Lemma \ref{detxi} and Lemma \ref{Dq2HS}.
\end{proof}

%%%%%%%%%%%%%%%%%%%%%%%%%%%%%%%%%%%%%%%%%%%%%%%%%%%%%%%%%%%%%%%%%%%%%%%%%%%%%%%%%%%%%%%%%%%%%%%%%%%%%%%

%%%%%%%%%%%%%%%%%%%%%%%%%%%%%%%%%%%%%%%%%%%%%%%%%%%%%%%%%%%%%%%%%%%%%%%%%%%%%%%%%%%%%%%%%%%%%%%%%%%%%%%

\section{Recurrence relation and difference equations}

\label{secpral}

%%%%%%%%%%%%%%%%%%%%%%%%%%%%%%%%%%%%%%%%%%%%%%%%%%%%%%%%%%%%%%%%%%%%%%%%%%%%%%%%%%%%%%%%%%%%%%%%%%%%%%%

%%%%%%%%%%%%%%%%%%%%%%%%%%%%%%%%%%%%%%%%%%%%%%%%%%%%%%%%%%%%%%%%%%%%%%%%%%%%%%%%%%%%%%%%%%%%%%%%%%%%%%%

As in the previous sections, we consider the Sobolev-type polynomials related to Charlier or Meixner families, maintaining the notation of the previous sections.

The main results of the first part of this work are stated in the present section. First, we provide in Theorem~\ref{S4-Theor3TRR-RC} a three-term recurrence relation for the Sobolev-type polynomials  $\{\mathbb{S}_{n}(x)\}_{n \geq 0}$, constructed in the previous section. We also describe two second order difference equations satisfied by these families of orthogonal polynomials (Theorem~\ref{sordDEqI} and Theorem~\ref{sordDEqII}).

%%%%%%%%%%%%%%%%%%%%%%%%%%%%%%%%%%%%%%%%%%%%%%%%%%%%%%%%%%%%%%%%%%%%%%%%%%%%%%%%%%%%%%%%%%%%%%%%%%%%%%%

\begin{theorem}
	\label{S4-Theor3TRR-RC} The following three-term recurrence relations hold for all $n\geq 1$, 
	\begin{equation*}
	\Xi _{2,n}(x)\mathbb{S}_{n+1}(x)=\overline{\alpha}_{n}(x)\mathbb{S}_{n}(x)+\overline{\beta}
	_{n}(x)\mathbb{S}_{n-1}(x),
	\end{equation*}%
	where 
	\begin{equation*}
	\Xi _{2,n}(x)=\Xi _{1,n}(x){\mathcal{E}}_{4,n+1}(x),
	\end{equation*}%
	\begin{equation*}
	\overline{\alpha} _{n}(x)=\Xi _{1,n+1}(x){\mathcal{E}}_{8,n}(x)-\Xi _{1,n}(x){\mathcal{F%
	}}_{4,n+1}(x),
	\end{equation*}%
	and 
	\begin{equation*}
	\overline{\beta} _{n}(x)=\Xi _{1,n+1}(x){\mathcal{F}}_{8,n}(x).
	\end{equation*}
\end{theorem}

%%%%%%%%%%%%%%%%%%%%%%%%%%%%%%%%%%%%%%%%%%%%%%%%%%%%%%%%%%%%%%%%%%%%%%%%%%%%%%%%%%%%%%%%%%%%%%%%%%%%%%%

%%%%%%%%%%%%%%%%%%%%%%%%%%%%%%%%%%%%%%%%%%%%%%%%%%%%%%%%%%%%%%%%%%%%%%%%%%%%%%%%%%%%%%%%%%%%%%%%%%%%%%%

\begin{proof}
	After shifting the index in (\ref{DqHsE3F3}) from $n$ to $n+1$ and using the
	recurrence relation (\ref{ReR}), one arrives at 
	\begin{equation*}
	\Delta\mathbb{S}_{n+1}(x)={\mathcal{E}}_{7,n}(x)P_{n}(x)+{\mathcal{F}}_{7,n}(x)P_{n-1}(x),
	\end{equation*}%
	where 
	\begin{equation*}
	{\mathcal{E}}_{7,n}(x)=(x-\alpha_{n}){\mathcal{E}}_{3,n+1}(x)+{\mathcal{F}}%
	_{3,n+1}(x),\quad \mbox{and}\quad {\mathcal{F}}_{7,n}(x)=-\beta _{n}{%
		\mathcal{E}}_{3,n+1}(x).
	\end{equation*}%
In view of Lemma \ref{detxi}, we deduce that 
	\begin{equation}
	\Xi _{1,n}(x)\Delta\mathbb{S}_{n+1}(x)={\mathcal{E}}_{8,n}(x)%
	\mathbb{S}_{n}(x)+{\mathcal{F}}_{8,n}(x)\mathbb{S}_{n-1}(x),
	\label{DqHsnm1}
	\end{equation}%
	where 
	\begin{equation*}
	{\mathcal{E}}_{8,n}(x)=-%
	\begin{vmatrix}
	{\mathcal{E}}_{2,n}(x) & {\mathcal{E}}_{7,n}(x) \\ 
	{\mathcal{F}}_{2,n}(x) & {\mathcal{F}}_{7,n}(x)%
	\end{vmatrix}%
	,
	\end{equation*}%
	and 
	\begin{equation*}
	{\mathcal{F}}_{8,n}(x)=%
	\begin{vmatrix}
	{\mathcal{E}}_{1,n}(x) & {\mathcal{E}}_{7,n}(x) \\ 
	{\mathcal{F}}_{1,n}(x) & {\mathcal{F}}_{7,n}(x)%
	\end{vmatrix}%
	.
	\end{equation*}%
	On the other hand, from Proposition \ref{STHST} we have%
	\begin{equation*}
	\Xi _{1,n+1}(x)\Xi _{1,n}(x)\Delta\mathbb{S}_{n+1}(x)=
	\Xi _{1,n}(x){\mathcal{E}}_{4,n+1}(x)\mathbb{S}_{n+1}(x)+\Xi _{1,n}(x){%
		\mathcal{F}}_{4,n+1}(x)\mathbb{S}_{n}(x).
	\end{equation*}
	Finally, substituting \eqref{DqHsnm1} in the previous expression, we conclude the result.
\end{proof}

%%%%%%%%%%%%%%%%%%%%%%%%%%%%%%%%%%%%%%%%%%%%%%%%%%%%%%%%%%%%%%%%%%%%%%%%%%%%%%%%%%%%%%%%%%%%%%%%%%%%%%%

%%%%%%%%%%%%%%%%%%%%%%%%%%%%%%%%%%%%%%%%%%%%%%%%%%%%%%%%%%%%%%%%%%%%%%%%%%%%%%%%%%%%%%%%%%%%%%%%%%%%%%%

\begin{theorem}[Second order difference equation, I]
	\label{sordDEqI} Let $\{\mathbb{S}_{n}(x)\}_{n \geq 0}$ be the sequence of Sobolev type polynomials. Then, the
	following statement holds.
	\begin{equation}
	{\mathcal{R}}_{n}(x)\Delta^{2}\mathbb{S}_{n}(x)+{\mathcal{S}}%
	_{n}(x)\Delta\mathbb{S}_{n}(x)+{\mathcal{T}}_{n}(x)\mathbb{S}_{n}(x)=0,\quad n \geq 0,  \label{SDEq1}
	\end{equation}
	where 
	\begin{equation*}
	{\mathcal{R}}_{n}(x)={\mathcal{F}}_{4,n}(x)\Xi_{1,n}(x),
	\end{equation*}
	\begin{equation*}
	{\mathcal{S}}_{n}(x) =-{\mathcal{F}}_{6,n}(x)\Xi_{1,n}(x),
	\end{equation*}
	and 
	\begin{equation*}
	{\mathcal{T}}_{n}(x) ={\mathcal{E}}_{4,n}(x){\mathcal{F}}_{6,n}(x)-{\mathcal{%
			E}}_{6,n}(x){\mathcal{F}}_{4,n}(x).
	\end{equation*}
\end{theorem}

%%%%%%%%%%%%%%%%%%%%%%%%%%%%%%%%%%%%%%%%%%%%%%%%%%%%%%%%%%%%%%%%%%%%%%%%%%%%%%%%%%%%%%%%%%%%%%%%%%%%%%%

%%%%%%%%%%%%%%%%%%%%%%%%%%%%%%%%%%%%%%%%%%%%%%%%%%%%%%%%%%%%%%%%%%%%%%%%%%%%%%%%%%%%%%%%%%%%%%%%%%%%%%%

\begin{proof}
	We have from Proposition \ref{STHST} that 
	\begin{equation}
	{\mathcal{F}}_{4,n}(x)\mathbb{S}_{n-1}(x)=\Xi _{1,n}(x)\Delta\mathbb{S}_{n}(x)-{\mathcal{E}}_{4,n}(x)\mathbb{S}_{n}(x).  \label{f4n}
	\end{equation}%
	The application of Proposition \ref{2SR} yields%
	\begin{equation*}
	{\mathcal{F}}_{4,n}(x)\Xi _{1,n}(x)\Delta^{2}\mathbb{S}_{n}(x)=
	{\mathcal{E}}_{6,n}(x){\mathcal{F}}_{4,n}(x)\mathbb{S}_{n}(x)+{\mathcal{F}}_{6,n}(x){\mathcal{F}}_{4,n}(x)\mathbb{S}_{n-1}(x).
	\end{equation*}%
	Then, from (\ref{f4n}) we get%
	\begin{equation*}
	{\mathcal{F}}_{4,n}(x)\Xi _{1,n}(x)\Delta^{2}\mathbb{S}_{n}(x)={\mathcal{E}}_{6,n}(x){\mathcal{F}}_{4,n}(x)\mathbb{S}_{n}(x)+{\mathcal{F}}_{6,n}(x)[\Xi _{1,n}(x)\Delta\mathbb{S}_{n}(x)-{\mathcal{E}}_{4,n}(x)\mathbb{S}_{n}(x)].
	\end{equation*}%
	We conclude that
	%
	%\begin{equation*}
	%{\mathcal{F}}_{4,n}(x)\Xi _{1,n}(x){\mathscr D}_{q}^{2}\mathbb{H}_{n}(x;q)-{%
	%\mathcal{E}}_{6,n}(x){\mathcal{F}}_{4,n}(x)\mathbb{H}_{n}(x;q)
	%\end{equation*}
	%\begin{equation*}
	%-{\mathcal{F}}_{6,n}(x)\Xi _{1,n}(x){\mathscr D}_{q}\mathbb{H}_{n}(x;q)+{%
	%\mathcal{E}}_{4,n}(x){\mathcal{F}}_{6,n}(x)\mathbb{H}_{n}(x;q)=0.
	%\end{equation*}%
	%Therefore, reagrouping the terms yields the conclusion
	\begin{equation*}
	{\mathcal{F}}_{4,n}(x)\Xi _{1,n}(x)\Delta^{2}\mathbb{S}_{n}(x)-{%
		\mathcal{F}}_{6,n}(x)\Xi _{1,n}(x)\Delta\mathbb{S}_{n}(x)
	+[{\mathcal{E}}_{4,n}(x){\mathcal{F}}_{6,n}(x)-{\mathcal{E}}_{6,n}(x){%
		\mathcal{F}}_{4,n}(x)]\mathbb{S}_{n}(x)=0.
	\end{equation*}
\end{proof}

%%%%%%%%%%%%%%%%%%%%%%%%%%%%%%%%%%%%%%%%%%%%%%%%%%%%%%%%%%%%%%%%%%%%%%%%%%%%%%%%%%%%%%%%%%%%%%%%%%%%%%%

%%%%%%%%%%%%%%%%%%%%%%%%%%%%%%%%%%%%%%%%%%%%%%%%%%%%%%%%%%%%%%%%%%%%%%%%%%%%%%%%%%%%%%%%%%%%%%%%%%%%%%%

%A similar analysis to that carried out in \cite{hermoso2020second} yields the following result, whose proof is sketched.

%%%%%%%%%%%%%%%%%%%%%%%%%%%%%%%%%%%%%%%%%%%%%%%%%%%%%%%%%%%%%%%%%%%%%%%%%%%%%%%%%%%%%%%%%%%%%%%%%%%%%%%

%%%%%%%%%%%%%%%%%%%%%%%%%%%%%%%%%%%%%%%%%%%%%%%%%%%%%%%%%%%%%%%%%%%%%%%%%%%%%%%%%%%%%%%%%%%%%%%%%%%%%%%

\begin{theorem}[Second order difference equation II]
	\label{sordDEqII} Let $\{\mathbb{S}_{n}(x)\}_{n\geq 0}$ be the Sobolev type polynomials. Then, the 	following statement holds for $n\geq 0$, 
	\begin{equation*}
	\overline{{\mathcal{R}}}_{n}(x)\Delta\nabla\mathbb{S}_{n}(x)+\overline{{\mathcal{S}}}_{n}(x)\nabla\mathbb{S}%
	_{n}(x)+\overline{{\mathcal{T}}}_{n}(x)\mathbb{S}_{n}(x)=0,\quad n\geq 0,
	\end{equation*}%
	where 
	\begin{equation*}
	\overline{{\mathcal{R}}}_{n}(x)={\mathcal{R}}_{n}(x-1),
	\end{equation*}%
	\begin{equation*}
	\overline{{\mathcal{S}}}_{n}(x)=[{\mathcal{S}}%
	_{n}(x-1)-{\mathcal{T}}_{n}(x-1)],
	\end{equation*}%
	and 
	\begin{equation*}
	\overline{{\mathcal{T}}}_{n}(x)={\mathcal{T}}_{n}(x-1).
	\end{equation*}
%	\begin{equation}
%	{\mathcal{R}}_{n}(x-1)\Delta\nabla\mathbb{S}_{n}(x)+[{\mathcal{S}}%
%	_{n}(x-1)-{\mathcal{T}}_{n}(x-1)]\nabla\mathbb{S}_{n}(x)+{\mathcal{T}}_{n}(x-1)\mathbb{S}_{n}(x)=0,\quad n \geq 0, 
%	\end{equation}
\end{theorem}

%%%%%%%%%%%%%%%%%%%%%%%%%%%%%%%%%%%%%%%%%%%%%%%%%%%%%%%%%%%%%%%%%%%%%%%%%%%%%%%%%%%%%%%%%%%%%%%%%%%%%%%

%%%%%%%%%%%%%%%%%%%%%%%%%%%%%%%%%%%%%%%%%%%%%%%%%%%%%%%%%%%%%%%%%%%%%%%%%%%%%%%%%%%%%%%%%%%%%%%%%%%%%%%

\begin{proof}
	It is a direct consequence of the property $\Delta f\left( x\right) =\nabla f\left(
	x+1\right) $ applied to \eqref{SDEq1} we have
	\begin{equation}
	{\mathcal{R}}_{n}(x)\Delta\nabla\mathbb{S}_{n}(x+1)+{\mathcal{S}}%
	_{n}(x)\nabla\mathbb{S}_{n}(x+1)+{\mathcal{T}}_{n}(x)\mathbb{S}_{n}(x)=0,\quad n \geq 0, 
	\end{equation}	
	Evaluating the resulting expression at $x-1$, the result follows.
%	\begin{equation}
%	{\mathcal{R}}_{n}(x-1)\Delta\nabla\mathbb{S}_{n}(x)+{\mathcal{S}}%
%	_{n}(x-1)\nabla\mathbb{S}_{n}(x)+{\mathcal{T}}_{n}(x-1)\mathbb{S}_{n}(x-1)=0,\quad n \geq 0, 
%	\end{equation}
%	Thus
%	\begin{equation}
%	{\mathcal{R}}_{n}(x-1)\Delta\nabla\mathbb{S}_{n}(x)+{\mathcal{S}}%
%	_{n}(x-1)\nabla\mathbb{S}_{n}(x)+{\mathcal{T}}_{n}(x-1)\mathbb{S}_{n}(x)-{\mathcal{T}}_{n}(x-1)\mathbb{S}_{n}(x)+{\mathcal{T}}_{n}(x-1)\mathbb{S}_{n}(x-1)=0,\quad n \geq 0, 
%	\end{equation}
%	\begin{equation}
%	{\mathcal{R}}_{n}(x-1)\Delta\nabla\mathbb{S}_{n}(x)+{\mathcal{S}}%
%	_{n}(x-1)\nabla\mathbb{S}_{n}(x)+{\mathcal{T}}_{n}(x-1)\mathbb{S}_{n}(x)-{\mathcal{T}}_{n}(x-1)[\mathbb{S}_{n}(x)-\mathbb{S}_{n}(x-1)]=0,\quad n \geq 0, 
%	\end{equation}
%	Hence,
%	\begin{equation}
%	{\mathcal{R}}_{n}(x-1)\Delta\nabla\mathbb{S}_{n}(x)+{\mathcal{S}}%
%	_{n}(x-1)\nabla\mathbb{S}_{n}(x)+{\mathcal{T}}_{n}(x-1)\mathbb{S}_{n}(x)-{\mathcal{T}}_{n}(x-1)\nabla\mathbb{S}_{n}(x)=0,\quad n \geq 0, 
%	\end{equation}
%	\begin{equation}
%	{\mathcal{R}}_{n}(x-1)\Delta\nabla\mathbb{S}_{n}(x)+[{\mathcal{S}}%
%	_{n}(x-1)-{\mathcal{T}}_{n}(x-1)]\nabla\mathbb{S}_{n}(x)+{\mathcal{T}}_{n}(x-1)\mathbb{S}_{n}(x)=0,\quad n \geq 0, 
%	\end{equation}
%	This allows to conclude the result.
\end{proof}

%%%%%%%%%%%%%%%%%%%%%%%%%%%%%%%%%%%%%%%%%%%%%%%%%%%%%%%%%%%%%%%%%%%%%%%%%%%%%%%%%%%%%%%%%%%%%%%%%%%%%%%

%%%%%%%%%%%%%%%%%%%%%%%%%%%%%%%%%%%%%%%%%%%%%%%%%%%%%%%%%%%%%%%%%%%%%%%%%%%%%%%%%%%%%%%%%%%%%%%%%%%%%%%

%%%%%%%%%%%%%%%%%%%%%%%%%%%%%%%%%%%%%%%%%%%%%%%%%%%%%%%%%%%%%%%%%%%%%%%%%%%%%%%%%%%%%%%%%%%%%%%%%%%%%%%

%%%%%%%%%%%%%%%%%%%%%%%%%%%%%%%%%%%%%%%%%%%%%%%%%%%%%%%%%%%%%%%%%%%%%%%%%%%%%%%%%%%%%%%%%%%%%%%%%%%%%%%
\section{Weighted Sobolev type polynomials}

\label{sec3}

%%%%%%%%%%%%%%%%%%%%%%%%%%%%%%%%%%%%%%%%%%%%%%%%%%%%%%%%%%%%%%%%%%%%%%%%%%%%%%%%%%%%%%%%%%%%%%%%%%%%%%%

%%%%%%%%%%%%%%%%%%%%%%%%%%%%%%%%%%%%%%%%%%%%%%%%%%%%%%%%%%%%%%%%%%%%%%%%%%%%%%%%%%%%%%%%%%%%%%%%%%%%%%%

This section is devoted to define the so-called weighted Sobolev
type polynomials, and describe their main properties, which will be used in
next Section in an application to watermarking schemes. The next result is a direct consequence of the definition of inner product. 

%%%%%%%%%%%%%%%%%%%%%%%%%%%%%%%%%%%%%%%%%%%%%%%%%%%%%%%%%%%%%%%%%%%%%%%%%%%%%%%%%%%%%%%%%%%%%%%%%%%%%%%

%%%%%%%%%%%%%%%%%%%%%%%%%%%%%%%%%%%%%%%%%%%%%%%%%%%%%%%%%%%%%%%%%%%%%%%%%%%%%%%%%%%%%%%%%%%%%%%%%%%%%%%

\begin{lemma}
	Let $\{\mathbb{S}_{n}(x)\}_{n\geq 0}$ be the Sobolev type polynomials. Then, the following statement holds for $n\geq 0$, 
	\begin{equation*}
		\lim\limits_{\lambda\rightarrow 0}||\mathbb{S}_{n}||_{\lambda}^2=||P_n||.
	\end{equation*}
\end{lemma}

\begin{definition}
	Let $\{\mathbb{S}_{n}(x)\}_{n\geq 0}$ be the sequence of Sobolev orthogonal polynomials. Then the weighted Sobolev type polynomial $\overline{\mathbb{S}}_{n}(x)$ is
	defined by
		\begin{equation}
		\overline{\mathbb{S}}_{n}(x)=\mathbb{S}_{n}(x)\sqrt{\frac{\rho(x)}{||\mathbb{S}_{n}||_{\lambda}^{2}}},  \label{wkrasob}
		\end{equation}
	for every $n\geq 0$.
\end{definition}

%%%%%%%%%%%%%%%%%%%%%%%%%%%%%%%%%%%%%%%%%%%%%%%%%%%%%%%%%%%%%%%%%%%%%%%%%%%%%%%%%%%%%%%%%%%%%%%%%%%%%%%

%%%%%%%%%%%%%%%%%%%%%%%%%%%%%%%%%%%%%%%%%%%%%%%%%%%%%%%%%%%%%%%%%%%%%%%%%%%%%%%%%%%%%%%%%%%%%%%%%%%%%%%

In the next result, we obtain the asymptotic behavior as $\lambda$ approaches to zero that leads to the asymptotic behavior of the matrix of orthogonal direct moments. The importance of orthogonal polynomials in watermarking schemes becomes clear from the next results, enunciated without proof that can be adapted from Lemma 3 and Proposition 5~\cite{HLS22}, and will also be clarified in Section~\ref{seck}.

%%%%%%%%%%%%%%%%%%%%%%%%%%%%%%%%%%%%%%%%%%%%%%%%%%%%%%%%%%%%%%%%%%%%%%%%%%%%%%%%%%%%%%%%%%%%%%%%%%%%%%%

%%%%%%%%%%%%%%%%%%%%%%%%%%%%%%%%%%%%%%%%%%%%%%%%%%%%%%%%%%%%%%%%%%%%%%%%%%%%%%%%%%%%%%%%%%%%%%%%%%%%%%%

\begin{lemma}
	\label{lemma_orthon_cond} Let $\{\mathbb{S}_{n}(x)\}_{n\geq 0}$ be the sequence of Sobolev orthogonal polynomials. Then, it holds that 
	\begin{equation*}
	\lim_{\lambda\rightarrow 0}\sum_{x\geq 0}\overline{\mathbb{S}}_{m}(x)\overline{\mathbb{S}}_{n}(x)=\delta _{m,n}.
	\end{equation*}
\end{lemma}

%%%%%%%%%%%%%%%%%%%%%%%%%%%%%%%%%%%%%%%%%%%%%%%%%%%%%%%%%%%%%%%%%%%%%%%%%%%%%%%%%%%%%%%%%%%%%%%%%%%%%%%

The following recurrence relation holds for the sequence of weighted Sobolev type polynomials.

%%%%%%%%%%%%%%%%%%%%%%%%%%%%%%%%%%%%%%%%%%%%%%%%%%%%%%%%%%%%%%%%%%%%%%%%%%%%%%%%%%%%%%%%%%%%%%%%%%%%%%%

%%%%%%%%%%%%%%%%%%%%%%%%%%%%%%%%%%%%%%%%%%%%%%%%%%%%%%%%%%%%%%%%%%%%%%%%%%%%%%%%%%%%%%%%%%%%%%%%%%%%%%%

\begin{proposition}
	\label{S3-Prop5} Let $\{\mathbb{S}_{n}(x)\}_{n\geq 0}$ be the sequence of Sobolev orthogonal polynomials. Then, the following recurrence relation holds: 
	\begin{equation*}
	\overline{\mathbb{S}}_{n+1}(x)=\Psi _{1,n}(x)\,\overline{\mathbb{S}}_{n}(x)+\Psi _{2,n}(x)\,\overline{\mathbb{S}}_{n-1}(x),\quad n\geq 0,
	\end{equation*}%
	where%
	\begin{equation*}
	\Psi _{1,n}(x)=\frac{||\mathbb{S}_{n}||_{\lambda}}{||\mathbb{S}_{n+1}||_{\lambda}}\,\frac{\overline{\alpha}_{n}(x)}{\Xi _{2,n}(x)}\quad \mbox{and}\quad \Psi _{2,n}(x)=\frac{||\mathbb{S}_{n-1}||_{\lambda}}{||\mathbb{S}_{n+1}||_{\lambda}}\,\frac{\overline{\beta}_{n}(x)}{\Xi _{2,n}(x)},
	\end{equation*}
	where $\Xi _{2,n}(x)$, $\overline{\alpha}_{n}(x)$ and $\overline{\beta}_{n}(x)$ is given by the Theorem \ref{S4-Theor3TRR-RC}.
\end{proposition}

%%%%%%%%%%%%%%%%%%%%%%%%%%%%%%%%%%%%%%%%%%%%%%%%%%%%%%%%%%%%%%%%%%%%%%%%%%%%%%%%%%%%%%%%%%%%%%%%%%%%%%%

\section{Watermarking scheme and related algorithms}\label{seck}

\label{sec4}

%%%%%%%%%%%%%%%%%%%%%%%%%%%%%%%%%%%%%%%%%%%%%%%%%%%%%%%%%%%%%%%%%%%%%%%%%%%%%%%%%%%%%%%%%%%%%%%%%%%%%%%

%%%%%%%%%%%%%%%%%%%%%%%%%%%%%%%%%%%%%%%%%%%%%%%%%%%%%%%%%%%%%%%%%%%%%%%%%%%%%%%%%%%%%%%%%%%%%%%%%%%%%%%

In this section, we preserve the notation of the previous sections, whose main concepts will be used to construct a watermarking scheme. In a first part, we briefly describe the watermarking process together with the embedding and extracting algorithms.

Let us consider an image, say $\mathcal{C}$, known as the cover image, and divide it into a certain number of smalled matrices of size $N\times N$ bytes. The $k-$%
th image block of $\mathcal{C}$ is denoted by ${\mathcal{C}}^{(k,N)}$. We denote
\begin{equation*}
{\mathcal{C}}^{(k,N)}=%
\begin{pmatrix}
{\mathcal{C}}_{0,0}^{(k,N)} & \cdots & {\mathcal{C}}_{0,N-1}^{(k,N)} \\ 
\vdots & \ddots & \vdots \\ 
{\mathcal{C}}_{N-1,0}^{(k,N)} & \cdots & {\mathcal{C}}_{N-1,N-1}^{(k,N)}%
\end{pmatrix}%
.
\end{equation*}

%%%%%%%%%%%%%%%%%%%%%%%%%%%%%%%%%%%%%%%%%%%%%%%%%%%%%%%%%%%%%%%%%%%%%%%%%%%%%%%%%%%%%%%%%%%%%%%%%%%%%%%

%%%%%%%%%%%%%%%%%%%%%%%%%%%%%%%%%%%%%%%%%%%%%%%%%%%%%%%%%%%%%%%%%%%%%%%%%%%%%%%%%%%%%%%%%%%%%%%%%%%%%%%

\begin{definition}
	Let ${\mathcal{C}}^{(k,N)}$ be the $k-$th image block of a cover image $%
	\mathcal{C}$. The associated matrix of orthogonal direct moments is given by
		\begin{equation}
		{\mathcal{M}} ={\mathcal{A}}\,{\mathcal{C}}^{(k,N)}{\mathcal{A}}^{t},
		\label{diretm}
		\end{equation}
	where
		\begin{equation}
		{\mathcal{A}}=%
		\begin{pmatrix}
		\overline{\mathbb{S}}_{0}(0) & \cdots & \overline{\mathbb{S}}%
		_{N-1}(0) \\ 
		\vdots & \ddots & \vdots \\ 
		\overline{\mathbb{S}}_{0}(N-1) & \cdots & \overline{\mathbb{S}}%
		_{N-1}(N-1)%
		\end{pmatrix}%
		.  \label{orthon}
		\end{equation}
\end{definition}

%%%%%%%%%%%%%%%%%%%%%%%%%%%%%%%%%%%%%%%%%%%%%%%%%%%%%%%%%%%%%%%%%%%%%%%%%%%%%%%%%%%%%%%%%%%%%%%%%%%%%%%

%%%%%%%%%%%%%%%%%%%%%%%%%%%%%%%%%%%%%%%%%%%%%%%%%%%%%%%%%%%%%%%%%%%%%%%%%%%%%%%%%%%%%%%%%%%%%%%%%%%%%%%

A steganographic image is an image created from the cover image which hides some additional data in it. Such additional data is subtle in such a way that the cover and the steganographic images remain close one to each other. In view of Lemma \ref{lemma_orthon_cond} a variation of the parameter $\lambda$ permits refined approximations of the image, due to ${\mathcal{A}}^{t}{\mathcal{A}}\approxeq{%
	\mathcal{I}}$, where ${\mathcal{I}}$ denote the identity matrix. We define the matrix of orthogonal inverse moments as follows.

%%%%%%%%%%%%%%%%%%%%%%%%%%%%%%%%%%%%%%%%%%%%%%%%%%%%%%%%%%%%%%%%%%%%%%%%%%%%%%%%%%%%%%%%%%%%%%%%%%%%%%%

%%%%%%%%%%%%%%%%%%%%%%%%%%%%%%%%%%%%%%%%%%%%%%%%%%%%%%%%%%%%%%%%%%%%%%%%%%%%%%%%%%%%%%%%%%%%%%%%%%%%%%%

\begin{definition}
	Let ${\mathcal{M}}$ be the matrix of orthogonal direct moments. Let $\mathcal{W}^{(k,N)}$ be defined by
		\begin{equation}
		\mathcal{W}^{(k,N)}={\mathcal{A}}^{t}\,{\mathcal{M}}{\mathcal{A}}.
		\label{inversem}
		\end{equation}
		We say $\mathcal{W}^{(k,N)}$ is the matrix of orthogonal inverse moments.
\end{definition}

%%%%%%%%%%%%%%%%%%%%%%%%%%%%%%%%%%%%%%%%%%%%%%%%%%%%%%%%%%%%%%%%%%%%%%%%%%%%%%%%%%%%%%%%%%%%%%%%%%%%%%%

%%%%%%%%%%%%%%%%%%%%%%%%%%%%%%%%%%%%%%%%%%%%%%%%%%%%%%%%%%%%%%%%%%%%%%%%%%%%%%%%%%%%%%%%%%%%%%%%%%%%%%%

We observe from Lemma \ref{lemma_orthon_cond} and (\ref{diretm}) that 
\begin{equation*}
\mathcal{W}^{(k,N)}={\mathcal{A}}^{t}\,{\mathcal{M}}{\mathcal{A}}={\mathcal{A%
}}^{t}\,{\mathcal{A}}\,{\mathcal{C}}^{(k,N)}{\mathcal{A}}^{t}{\mathcal{A}}
\end{equation*}%
which entails %
\begin{equation*}
\lim_{\lambda\rightarrow 0}\mathcal{W}^{(k,N)}={\mathcal{C}}%
^{(k,N)}.
\end{equation*}

The variation of $\lambda$ allows to maintain as much similarity as wanted between the cover and the steganographic image. The procedure to change from one to another image is done via the embedding algorithm, providing a watermarked image $\mathcal{W}$ from a cover image $\mathcal{C}$. Before introducing this and the extracting algorithm which provides with a robust watermark from a watermarked image, we clarify the way to produce a pixel scrambling.

\subsection{Piecewise linear chaotic map}

In this work, the piecewise linear chaotic map (PWLCM) is used for pixel scrambling. This technique is given by
\begin{eqnarray*}
	x_{i+1}&=&F(x_i,\mu)\\
	&=&\displaystyle\begin{cases}
		\displaystyle\frac{x_i}{\mu}, & \mbox{if }x_i\in (0,\mu]\\\\
			\displaystyle\frac{x_i-\mu}{0\mbox{.}5-\mu}, & \mbox{if }x_i\in(\mu,0\mbox{.}5]\\\\
			F(1-x_i,\mu), & \mbox{if }x_i\in(0\mbox{.}5,1)		
	\end{cases}
\end{eqnarray*}
where $x_i\in(0, 1)$. Taking the control parameter $\mu\in(0, 0{.}5)$, it evolves into a chaotic state. Then, the chaotic permutation of $\varrho=\{\varrho_1,\ldots,\varrho_n\}$ used for pixel scrambling is determined by $(\varrho_j)_{j\in{\mathcal{P}}}$ where
\begin{equation*}
{\mathcal{P}}=\left\lbrace \floor*{x_k10^{14}\mbox{ mod. }n},\quad 1\leq k\leq n\right\rbrace , \label{fcmg}
\end{equation*}
being $n$ the number of pixels. Here, the pair $(x,\mu)$ is used as a secret key.

\begin{figure}[h!]
	\centering\includegraphics[width=0.2\textwidth]{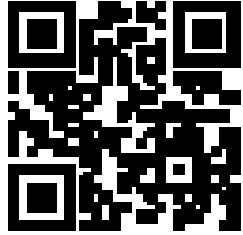}
	\caption{Watermark of $64\times 64$ bits}
	\label{waterm}
\end{figure}

The watermark that will be used in this contribution is that of Figure~\ref{waterm}.

In the following section, we write two algorithms involved in the watermarking procedure. More precisely, the embedding and the extracting algorithms, providing a watermarked image from a cover image, and viceversa. In these algorithms, we make use of the zigzag scan and Dither Modulation which are described in detail in~\cite{Soria-2022} and~\cite{HLS22}, respectively. We have omitted them for the sake of readability in a more compact writing.

\subsection{Embedding and extraction watermark algorithm}

%Firstly, the embedding watermark algorithm scrambled watermark by the Arnold transform \eqref{arTransform} for a control parameter $\eta$, con $1\leq \eta\leq 63$, see Figure \ref{arnold}, and then organized into a binary sequence $\{\hat{\omega}_i\}$, $i=1,\ldots, 64\times 64$. Next, this scheme splits the cover image $\mathcal{C}$ up into non-overlapping blocks ${\mathcal{C}}^{(k,8)}$ of size $8\times 8$, where the number of blocks coincide the number of watermark bits. And the other hand, it applies the direct moments \eqref{diretm} to each block of $8\times 8$. Next, the zigzag scan is applied to the resultant coefficients block, see Figure \ref{zigzag}, with the purpose to align frequency coefficients in ascending order. Then, it selects a coefficient of \eqref{diretm}, in this case, the coefficient number is 28. Thus, the secrete bits are embedded in the selected coefficient by applying the Dither Modulation (DM). Finally, the inverse moment transform \eqref{inversem} is applied in order to reconstruct the image, obtaining the watermarked image $\mathcal{W}^{(k,8)}$.
%
%Algorithm \ref{alg:embedding} describes the procedure explained in Section %
%\ref{sec4} at the time of watermarking a cover image $\mathcal{C}$.

\begin{algorithm}[H]
	\caption{Embedding Algorithm}  \label{alg:embedding}  
	\begin{algorithmic}[1]	
		\STATE \text{{\bf Input:}} Cover image $\mathcal{C}$, robust watermark $\{\omega_i\in\{0,1\}\mbox{ : }i=1,2,\ldots\}$, key $\kappa$, $x_0$, $\mu$.	
		
		\STATE \text{{\bf Output:}} Watermarked image ${\mathcal{W}}$
		
		\STATE \text{{\bf Fragile watermark:}} $\upsilon\gets\textbf{sha256}(\kappa)[:16]$ 
		
		\STATE $\{\hat{\omega}_i\}\gets$ scrambled watermark by the piecewise linear chaotic map $(x_0,\mu)$
		
		\STATE Divide ${\mathcal{C}}$ into non-overlapping blocks of $8\times8$ bytes
		
		%		\STATE $\ell=0$
		
		\FOR{\textbf{each} ${\mathcal C}^{(k,8)}\in{\mathcal{C}}$}
		
		\STATE ${\mathcal M}\gets{\mathcal A}\,{\mathcal C}^{(k,8)}{\mathcal A}^{t}$ : according to \eqref{diretm}
		
		\STATE $\nu^{k}\gets{\mathscr Z}({\mathcal M})$ : Apply the zigzag scan \cite[Section 4.3]{Soria-2022}
		
		\STATE $\overline{\nu}_{28}^{k}\gets\nu_{28}^{k}$ watermark bit $\hat{\omega}_k$ is embedded in the selected coefficient $\nu_{28}^{k}$ by using Dither Modulation, \cite[Section 5.3]{HLS22}
		
		%		\STATE $\vartheta\gets\{\nu_j^{k}\}_{2\leq j\leq 9} $
		%		
		%		\FOR{\textbf{each} $\vartheta_j\in\vartheta$}
		%		
		%		\STATE $\ell\gets\ell + 1$
		%		
		%		\IF{$\vartheta_j<0$}
		%		
		%		\STATE $\bar{\vartheta}_j\gets -$R$(\mbox{abs}(\vartheta_j), \omega_{\ell})$;
		%		
		%		\ELSE
		%		
		%		\STATE $\bar{\vartheta}_j\gets $R$(\vartheta_j, m_{\ell})$;
		%		
		%		\ENDIF
		%		
		%		\ENDFOR
		%		
		%		\STATE $\{\nu_j^{k}\}_{2\leq j\leq 9}\gets\bar{\vartheta}$
		
		\STATE $\overline{{\mathcal M}}\gets{\mathscr Z}^{-1}(\overline{\nu}^k)$ \cite[Section 4.3]{Soria-2022}
		
		\STATE $\overline{{\mathcal W}}^{(k,8)}\gets{\mathcal A}^{t}\,\overline{{\mathcal M}}\,{\mathcal A}$ : According to \eqref{inversem}
		
		\ENDFOR\\
		
		\FOR{\textbf{each} $\overline{{\mathcal W}}^{(k,8)}\in\overline{{\mathcal{W}}}$}
		
		\STATE $\varsigma\gets{\mathscr Z}(\overline{{\mathcal W}}^{(k,8)})$ : Apply the zigzag scan
		
		\STATE $\textbf{LSB}(\varsigma[:16])\gets\upsilon$: according to \cite[Section 4.3]{Soria-2022}
		
		\STATE ${\mathcal W}^{(k,8)}\gets{\mathscr Z}^{-1}(\varsigma)$ : According to \cite[Section 4.3]{Soria-2022}
		
		\ENDFOR\\
		
		\RETURN ${\mathcal{W}}$
	\end{algorithmic} 
\end{algorithm}

\begin{algorithm}[h]
	\caption{Extracting Algorithm}  \label{alg:extracting}  
	\begin{algorithmic}[1]	
		\STATE \text{{\bf Input:}} Watermarked image $\mathcal{W}$, key $\kappa$, $x_0$, $\mu$.	
		
		\STATE \text{{\bf Output:}} Robust watermark $\{\omega_i\in\{0,1\}\mbox{ : }i=1,2,\ldots\}$ and tamper detection
		
%		\STATE  $\upsilon\gets\textbf{sha256}(\kappa)[:16]$ 
		
%		\STATE $\{\hat{\omega}_i\}\gets$ scrambled watermark by the Arnold transform \eqref{arTransform}.
		
		\STATE Divide ${\mathcal{W}}$ into non-overlapping blocks of $8\times8$ bytes
		
		%		\STATE $\ell=0$
		
		\FOR{\textbf{each} ${\mathcal C}^{(k,8)}\in{\mathcal{C}}$}
		
		\STATE ${\mathcal M}\gets{\mathcal A}\,{\mathcal C}^{(k,8)}{\mathcal A}^{t}$ : according to \eqref{diretm}
		
		\STATE $\nu^{k}\gets{\mathscr Z}({\mathcal M})$ : Apply the zigzag scan
		
		\STATE $\omega_k\gets$ watermark bit is extracted from the selected coefficient $\nu_{28}^{k}$ by using Dither Modulation.
		
		%		\STATE $\vartheta\gets\{\nu_j^{k}\}_{2\leq j\leq 9} $
		%		
		%		\FOR{\textbf{each} $\vartheta_j\in\vartheta$}
		%		
		%		\STATE $\ell\gets\ell + 1$
		%		
		%		\IF{$\vartheta_j<0$}
		%		
		%		\STATE $\bar{\vartheta}_j\gets -$R$(\mbox{abs}(\vartheta_j), \omega_{\ell})$;
		%		
		%		\ELSE
		%		
		%		\STATE $\bar{\vartheta}_j\gets $R$(\vartheta_j, m_{\ell})$;
		%		
		%		\ENDIF
		%		
		%		\ENDFOR
		%		
		%		\STATE $\{\nu_j^{k}\}_{2\leq j\leq 9}\gets\bar{\vartheta}$
		
%		\STATE $\overline{{\mathcal M}}\gets{\mathscr Z}^{-1}(\overline{\nu}^k)$
%		
%		\STATE $\overline{{\mathcal W}}^{(k,8)}\gets{\mathcal A}^{t}\,\overline{{\mathcal M}}{\mathcal A}$ : According to \eqref{inversem}
		
		\ENDFOR\\
		
		\FOR{\textbf{each} $\overline{{\mathcal W}}^{(k,8)}\in\overline{{\mathcal{W}}}$}
		
		\STATE $\varsigma\gets{\mathscr Z}(\overline{{\mathcal W}}^{(k,8)})$ : Apply the zigzag scan
		
		\STATE $\upsilon\gets\textbf{LSB}(\varsigma[:16])$ \text{{\bf fragile watermark}}
		
		\IF{ $\upsilon\neq\textbf{sha256}(\kappa)[:16]$}
		
		\STATE $\mathcal{W}$ is not authentic; break

		\ENDIF
		
		\ENDFOR
	\end{algorithmic} 
\end{algorithm}

%%%%%%%%%%%%%%%%%%%%%%%%%%%%%%%%%%%%%%%%%%%%%%%%%%%%%%%%%%%%%%%%%%%%%%%%%%%%%%%%%%%%%%%%%%%%%%%%%%%%%%%

%%%%%%%%%%%%%%%%%%%%%%%%%%%%%%%%%%%%%%%%%%%%%%%%%%%%%%%%%%%%%%%%%%%%%%%%%%%%%%%%%%%%%%%%%%%%%%%%%%%%%%%
%%%%%%%%%%%%%%%%%%%%%%%%%%%%%%%%%%%%%%%%%%%%%%%%%%%%%%%%%%%%%%%%%%%%%%%%%%%%%%%%%%%%%%%%%%%%%%%%%%%%%%%

\section{Results and discussion}

The experimental results of the proposed scheme are now described through an experimental analysis of four images. The algorithm is implemented in Python 3.8.10.

For the experimental analysis 24 color images of size (512$\times$512) were collected from two different datasets: 

12 images of the image dataset of 1500 RGB-BMP images, transformed from Caltech birds' dataset in JPEGC format~\cite{AlJarrahM},

12 images of the image dataset of 1500 RGB-BMP images, transformed from NRC dataset in TIFF format~\cite{AlJarrahM}.

The experimental analysis reveal the values of PSNR (Peak
Signal to Noise Ratio \cite{Soria-2022}) and BER (bit error rate \cite{Soria-2022}) of Charlier-Sobolev type polynomials, Meixner-Sobolev type polyonomials, compared with the Krawtchouk-Sobolev type polynomials proposed by \cite{Soria-2022}. The notation used for the different experiments is as follows: the proposed method for Charlier-Sobolev type orthogonal moments is denoted by CS and comprises two choices of the parameter involved.

CS$\_$I: $\mu = 0.0007$, $\lambda = 10 ^ {-47}$, $\alpha = -17$, $j = 5$, 

CS$\_$II: $\mu = 0.0005$, $\lambda = 10 ^ {-77}$, $\alpha = -21$, $j = 3$.

On the other hand, Meixner-Sobolev type orthogonal moments will be denoted by MS, considering two choices for the values of the parameters involved in the construction of the family of orthogonal polynomials.

MS$\_$I: $\mu = 0.0008$, $\gamma = 0.000041$, $\lambda = 10^{-47}$, $\alpha = -17$, $j = 5$, 

MS$\_$II: $\mu = 0.0001$, $\gamma = 0.000075$, $\lambda = 10^{-77}$, $\alpha = -21$, $j = 3$. 

They are compared with the method proposed by \cite{Soria-2022} for Krawtchouk-Sobolev type orthogonal moments, denoted by $KS$. 

KS: $p=0.48$, $N=8$, $\lambda=10^{-77}$,  $j = 5$,

which have been proved to be efficient in a watermarking scheme. 

We have considered different attacks to measure robustness, namely Cropping noise, Gaussian noise, Gaussian Laplace, Minimum filter noise and Salt $\&$ Pepper noise. 

\begin{figure}[H] % [ht]
	\centering
	\begin{tabular}{cc}
		\includegraphics[scale=.47]{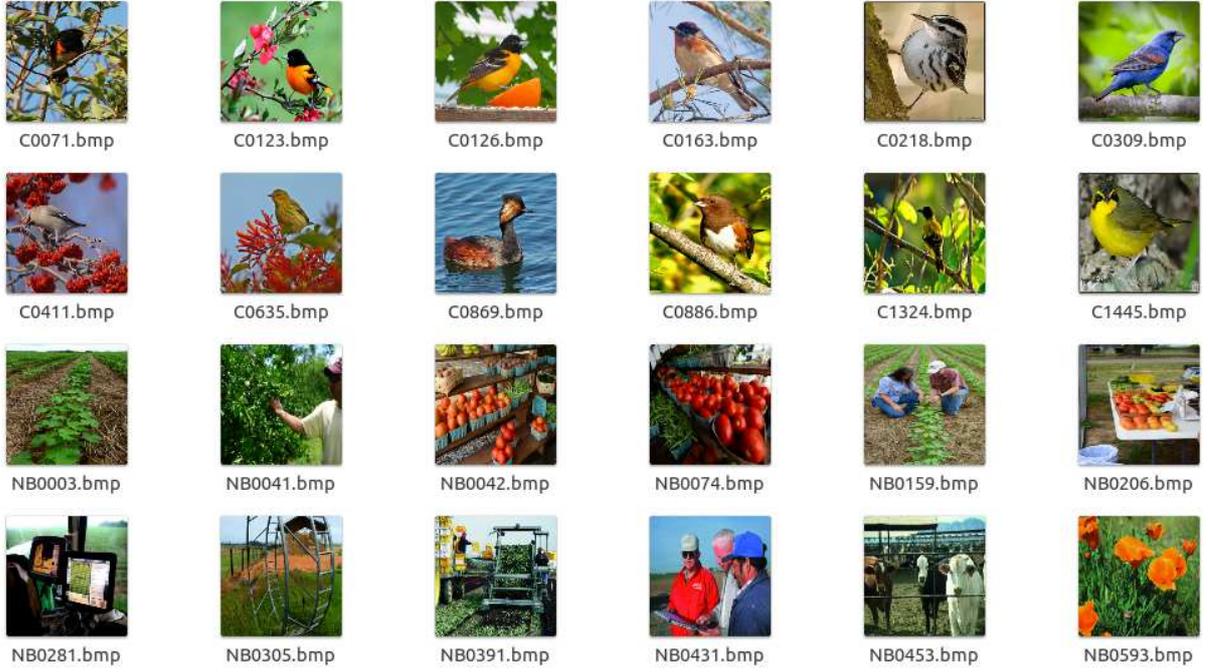}
	\end{tabular}
	\caption{Cover images}
	\label{covs}	
\end{figure}

\subsection{Imperceptibility test}
\begin{figure}[H] % [ht]
	\centering
	\includegraphics[scale=0.85]{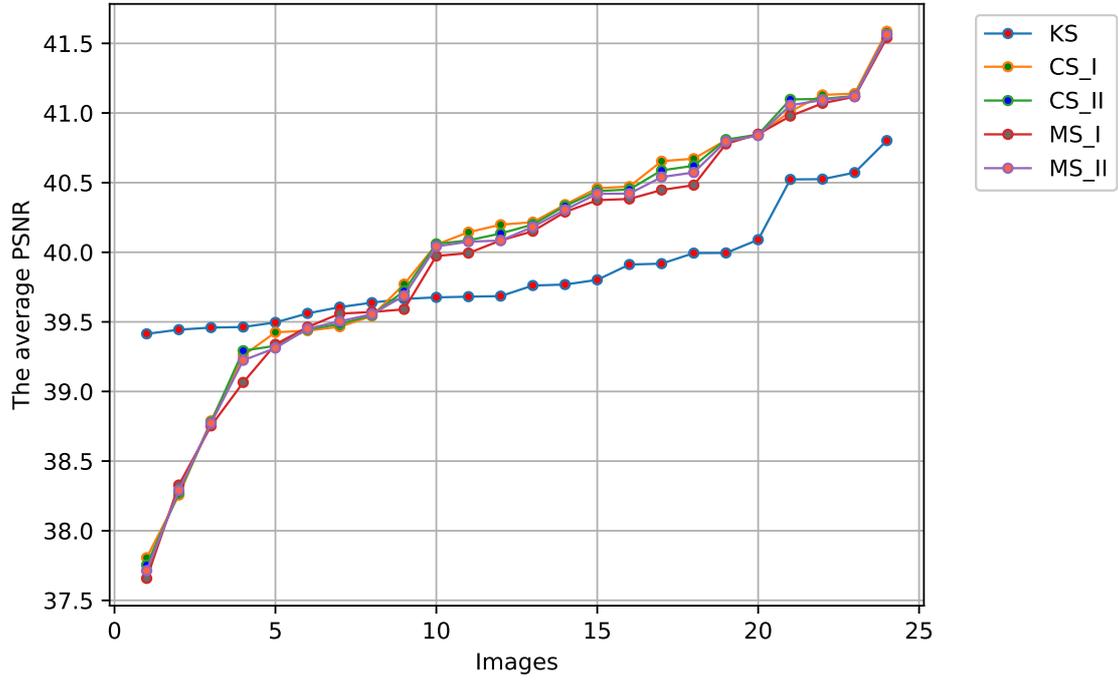}
	\caption{PSNR values}
	\label{psnrValues}	
\end{figure}

The experimental results showed that the proposed studied moments produced good quality watermarked documents with good PSNR values, between 37 and 42 db, which is in correspondence with the heuristic values of PSNR, see Figure \ref{psnrValues}. Indeed, the PSNR to evaluate the level of imperceptibility and distortion as well as to measure the difference between cover and watermarked documents. 

We recall that PSNR is given by 
\begin{equation*}
\text{PSNR}=10\log _{10}\left( \displaystyle{\frac{{\displaystyle{\Xi ^{2}}}%
}{{\displaystyle{\text{MSE}}}}}\right) ,
\end{equation*}%
with $\text{MSE}=(N^{2}\rho )^{-1}\sum_{\mathbf{\gamma }\in \Gamma }\left\Vert 
\mathcal{C}\left( \mathbf{\gamma }\right) -\mathcal{W}\left( \mathbf{\gamma }%
\right) \right\Vert ^{2}$.
$\mathcal{C},\mathcal{W}\in \{0,1,\dots ,\Xi \}$, and $\Xi =\text{max}(\text{max}(\mathcal{C}),\text{max}(\mathcal{W}))$, where $\mathcal{C}$ is the cover image and $\mathcal{W}$ represents the stego image, respectively, of size $N^{2}\rho $.

The index set ${\mathbf{\gamma}} = (\ell_1, \ell_2, \ell_3)$ is defined on the
set ${\Gamma} = \{1, \dots, m \} \times \{ 1, \dots, n \} \times \{1, \dots, \rho
\},$ for $\rho=1$ for gray scale images and $\rho=3$ for 24-bit color images.

Indeed, this experiment shows that for 24 images of the two datasets, the results of imperceptibility corresponding to the CS and MS methods are best for the KS method, see Figure~\ref{psnrValues}.

%{\color{red}{seria bueno incluir alguna grafica que comparara este parametro en los distintos experimentos, como la figura 5 de nuestro articulo en mathematics}}

\subsection{Robustness test}

Incorrectly formed binary values of the watermark image determine the robustness of the methodin terms of the bit error rate (BER). The BER value is computed by
\begin{equation*}
\mbox{BER}= \frac{1}{L}\sum_{n=0}^{L-1}\begin{cases}
1, & \overline{\omega}(n)\neq\omega(n),\\ 
0, & \overline{\omega}(n)=\omega(n),
\end{cases}
\end{equation*}
where $\omega(n)$ and $\overline{\omega}(n)$ are binary bits (0 or 1) of $\mathcal{C}$ and $\mathcal{W}$. Here, $L$ stands for the number of bits of the watermarking scheme.

In order to evaluate the robustness, the following attacks were applied, Cropping noise, Gaussian noise, Gaussian Laplace, Minimum filter noise and Salt $\&$ Pepper noise, where their parameters appear in Table \ref{noisepara}. 

\begin{table}[ht!]
	\caption{Information of the attacks applied}
	\label{noisepara}%[ht!]
	% tbph]
	\centering
	\renewcommand{\arraystretch}{2.0} 
	\begin{tabular}{lc}
\\ \hline
		Attacks (noises) & Parameters \\ \hline
		Cropping noise & Image percentage: $5\,\%$, $10\,\%$, $15\,\%$, $20\,\%$, $25\,\%$, $30\,\%$, $35\,\%$, $40\,\%$\\
		Fourier ellipsoid filter & Sizes of the box used for filtering: 1, 2, 3, 4, 5, 6, 7, 8\\
		Gaussian & Sigma of the Gaussian kernel: 0.1, 0.2, 0.3, 0.4, 0.5, 0.6, 0.7, 0.8\\
		Gaussian Laplace & Sigma of the Gaussian kernel: 0.01, 0.02, 0.03, 0.04, 0.05, 0.06, 0.07, 0.08\\
		Minimum filter & Kernel size: 1, 2, 3, 4, 5, 6, 7, 8\\
		Salt $\&$ Pepper & Density: 0.01, 0.02, 0.03, 0.04, 0.05, 0.06, 0.07, 0.08\\ \hline
	\end{tabular}%
\end{table}

\begin{figure}[H] % [ht]
	\centering
	\begin{tabular}{cc}
		\includegraphics[scale=.45]{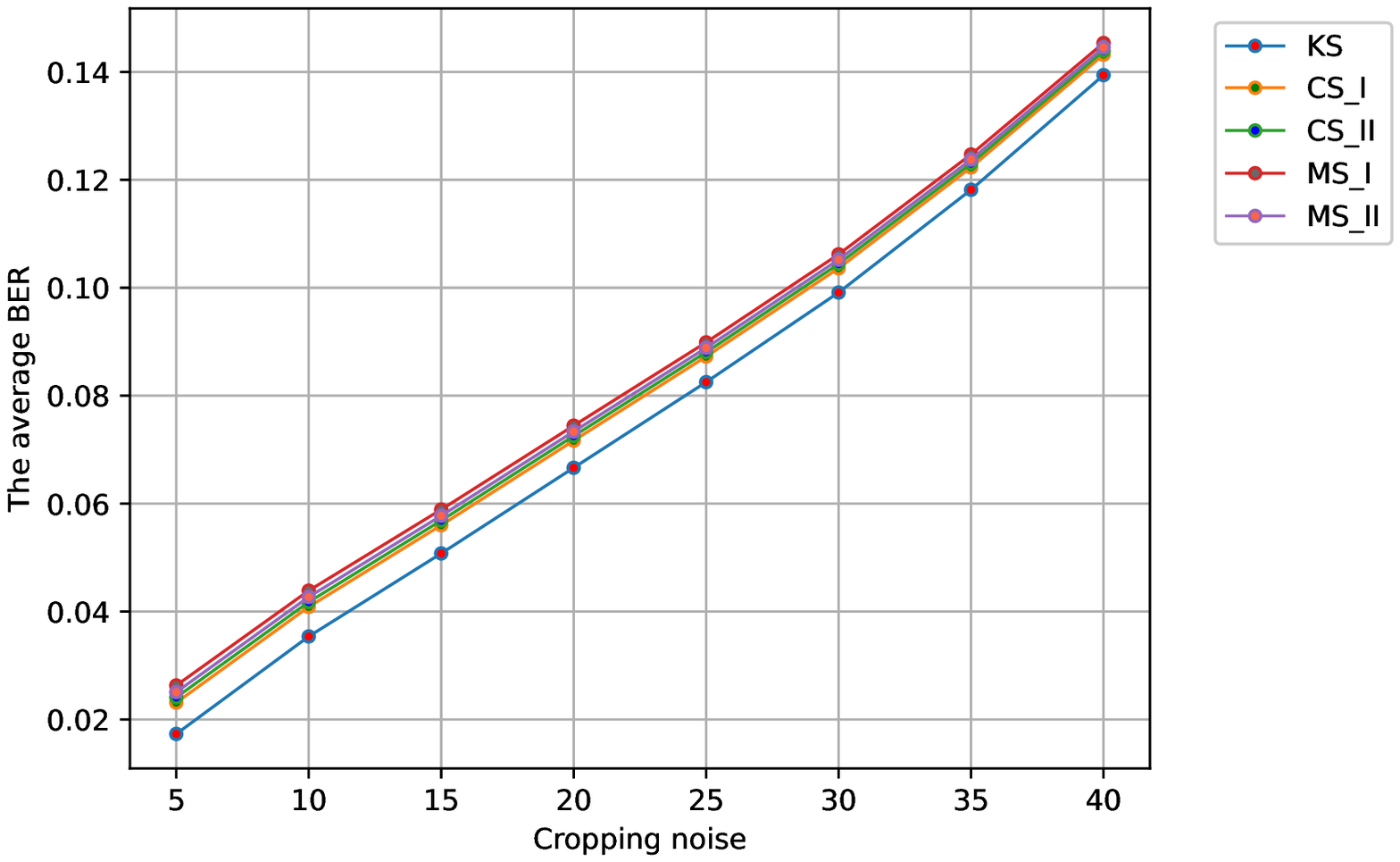}
		&
		\includegraphics[scale=.45]{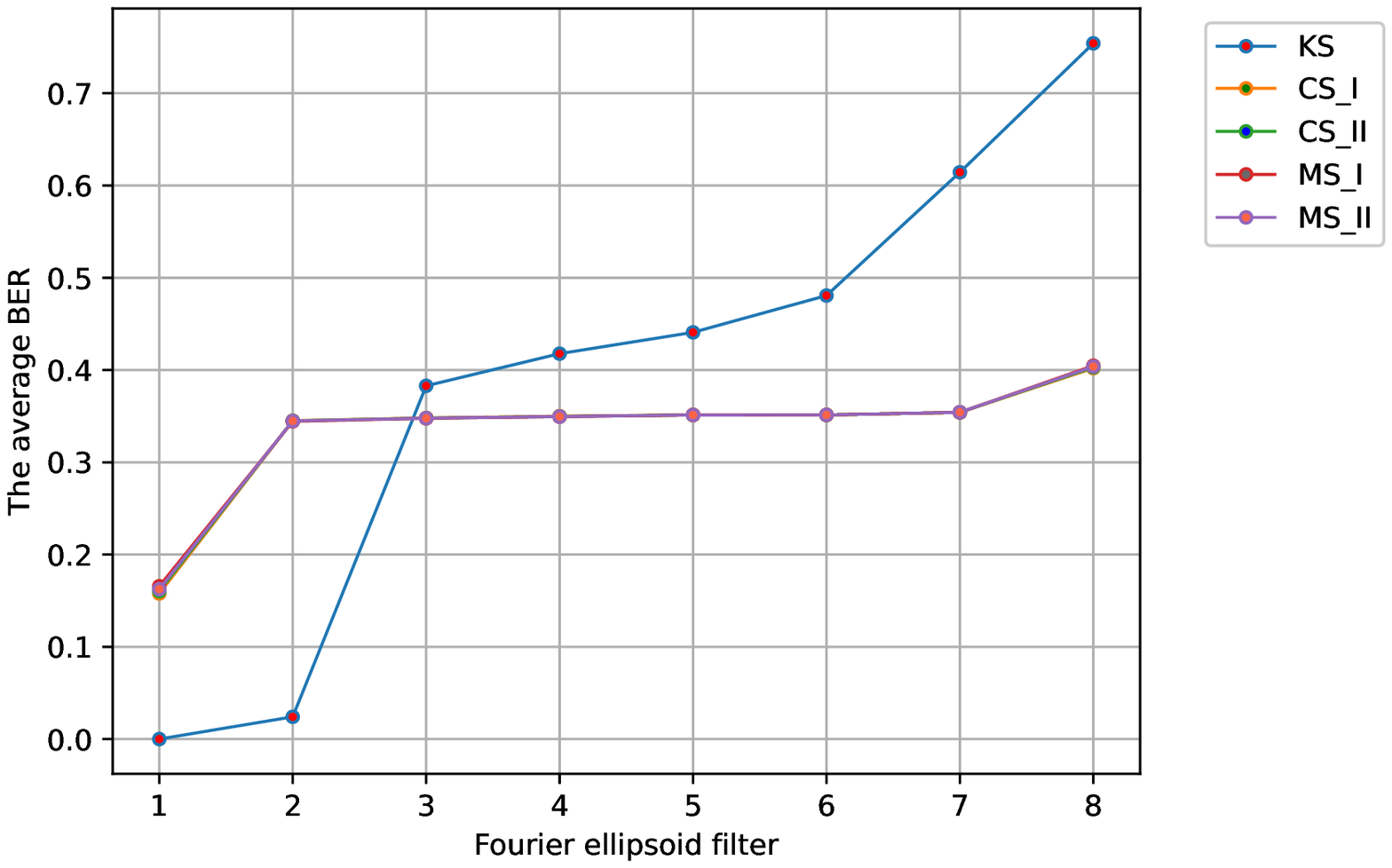}
		\\
		\includegraphics[scale=.45]{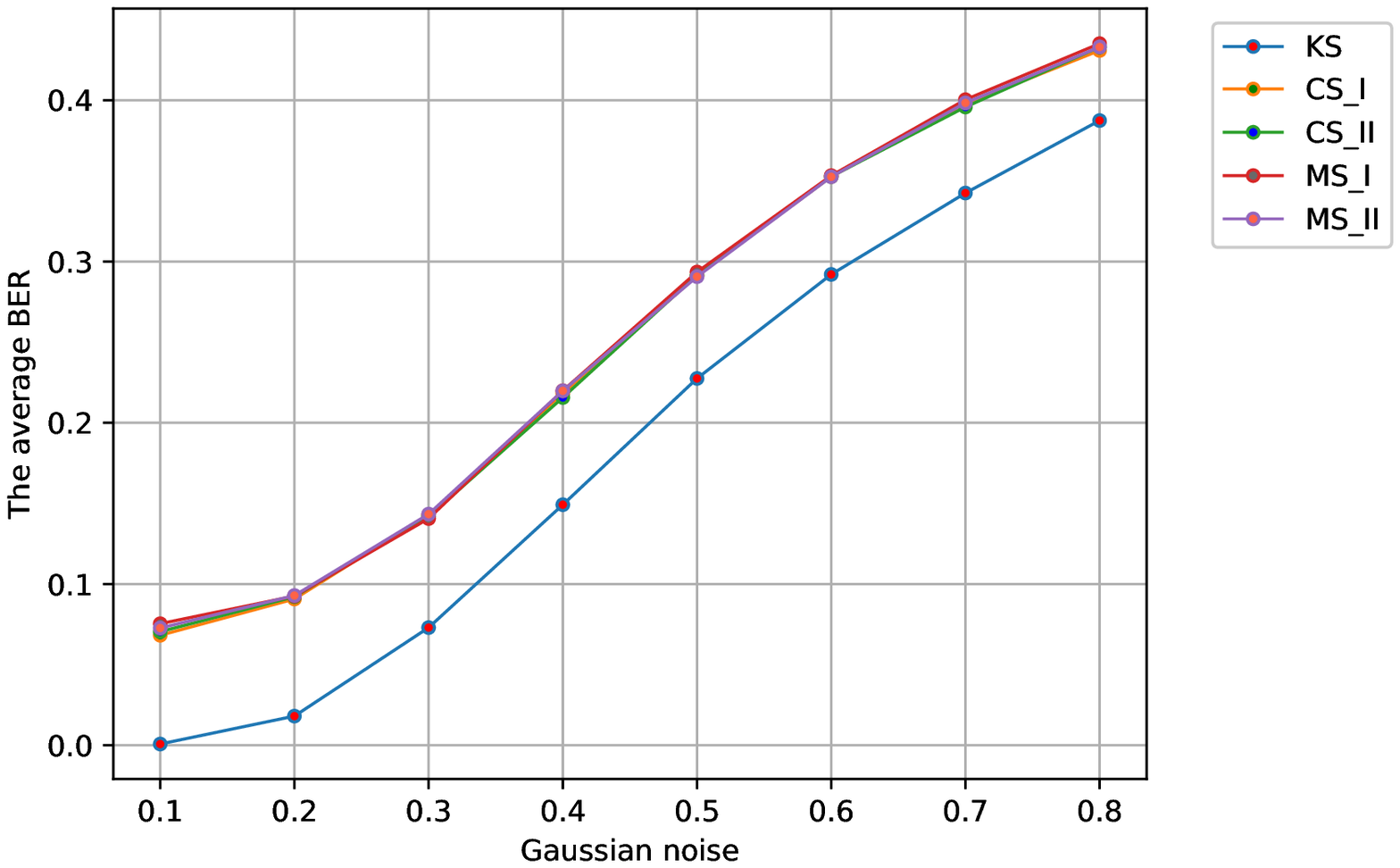}
		&
		\includegraphics[scale=.45]{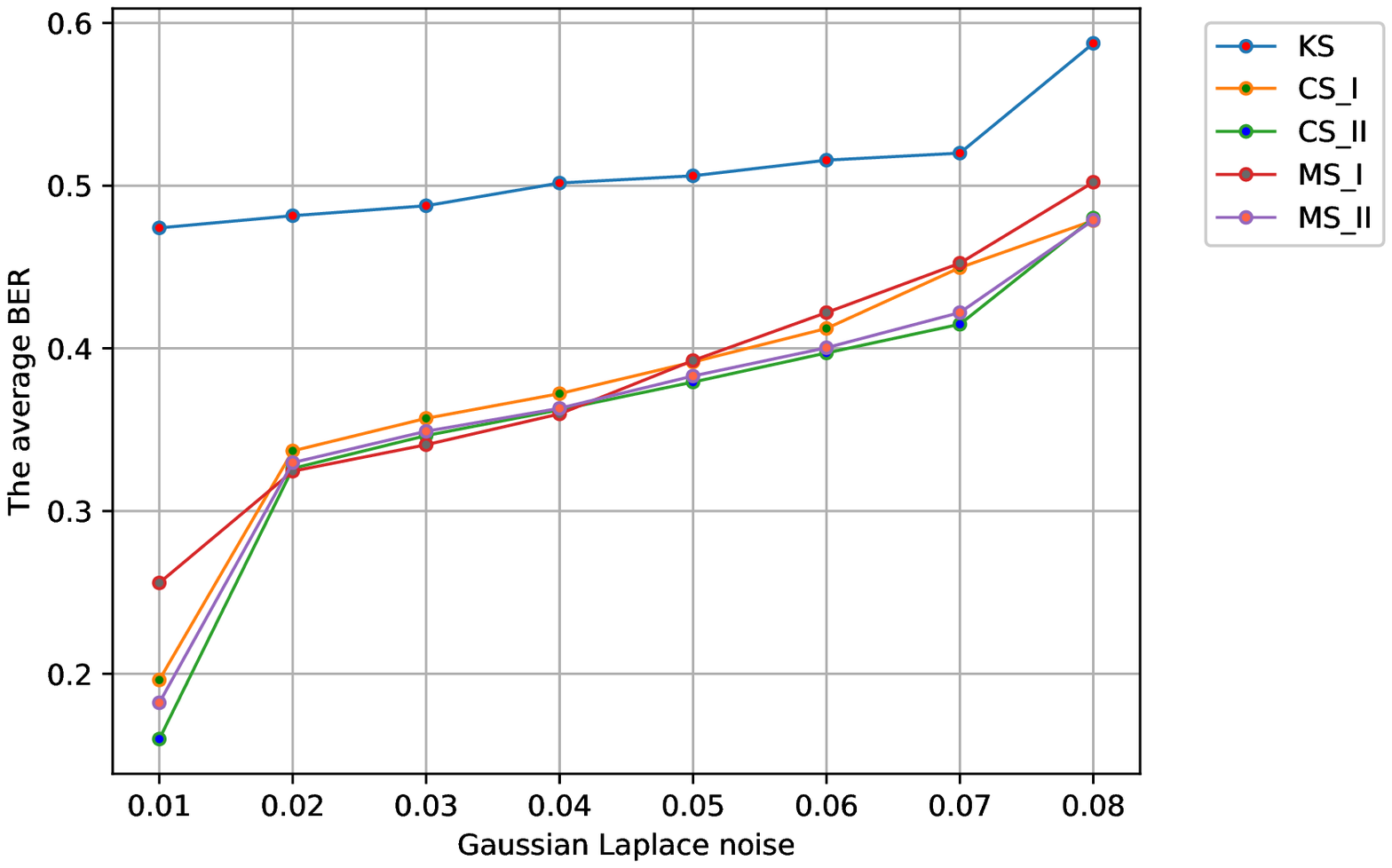}
		\\
		\includegraphics[scale=.45]{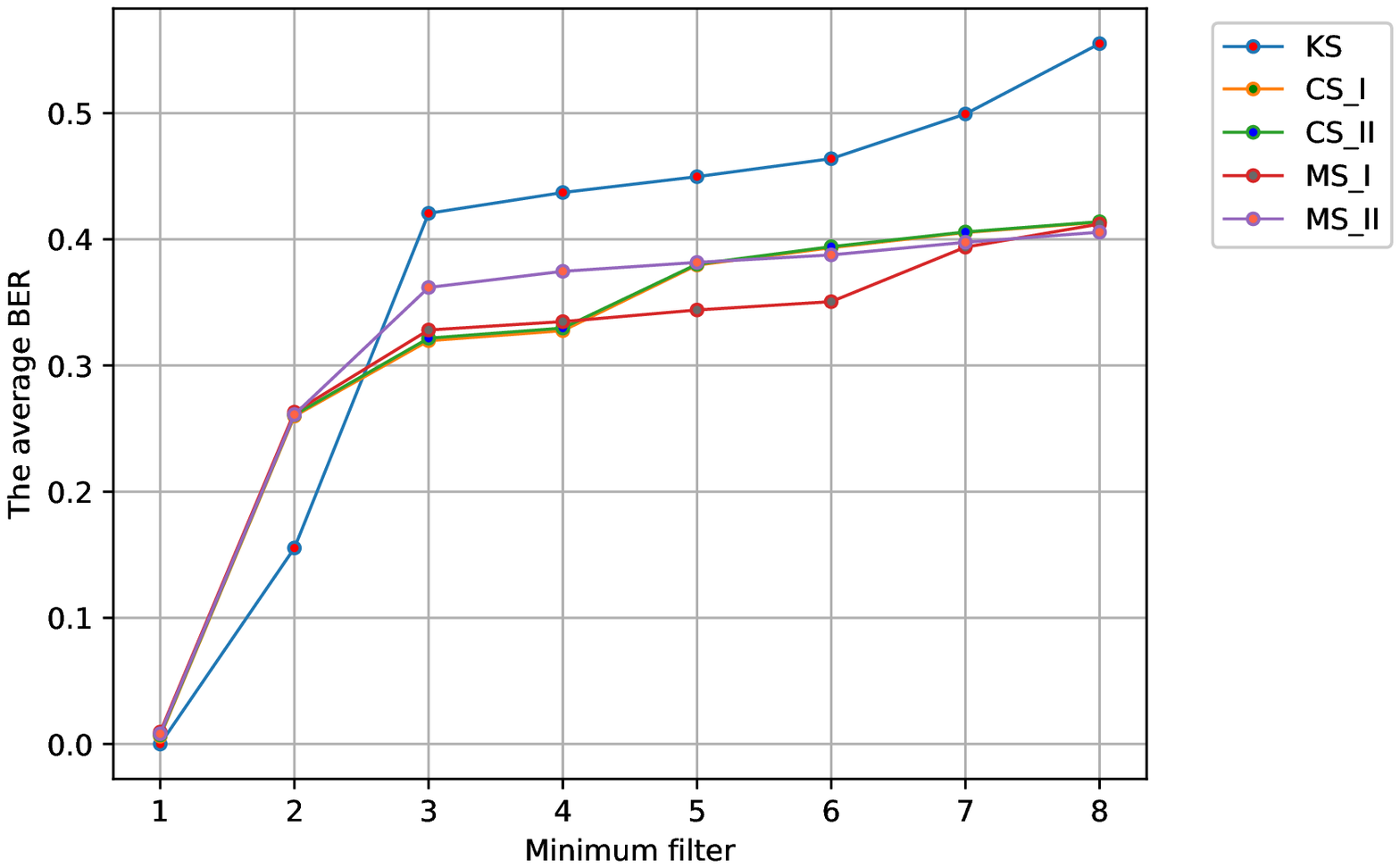}
		&
		\includegraphics[scale=.45]{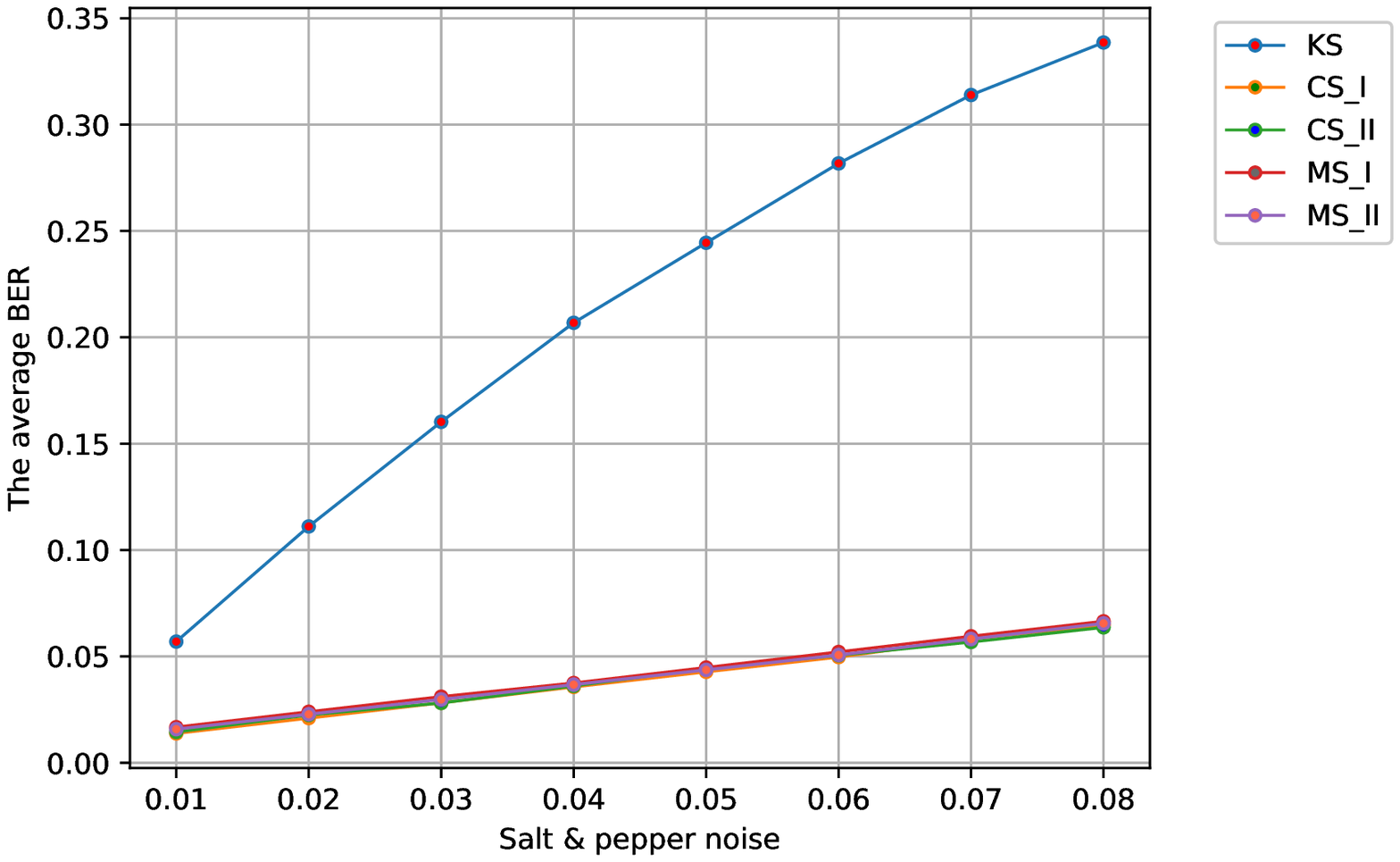}
	\end{tabular}
	\caption{Attacks applied}
	\label{attacks}	
\end{figure}

\begin{figure}[H] % [ht]
	\centering
	\begin{tabular}{cc}
		\includegraphics[scale=.44]{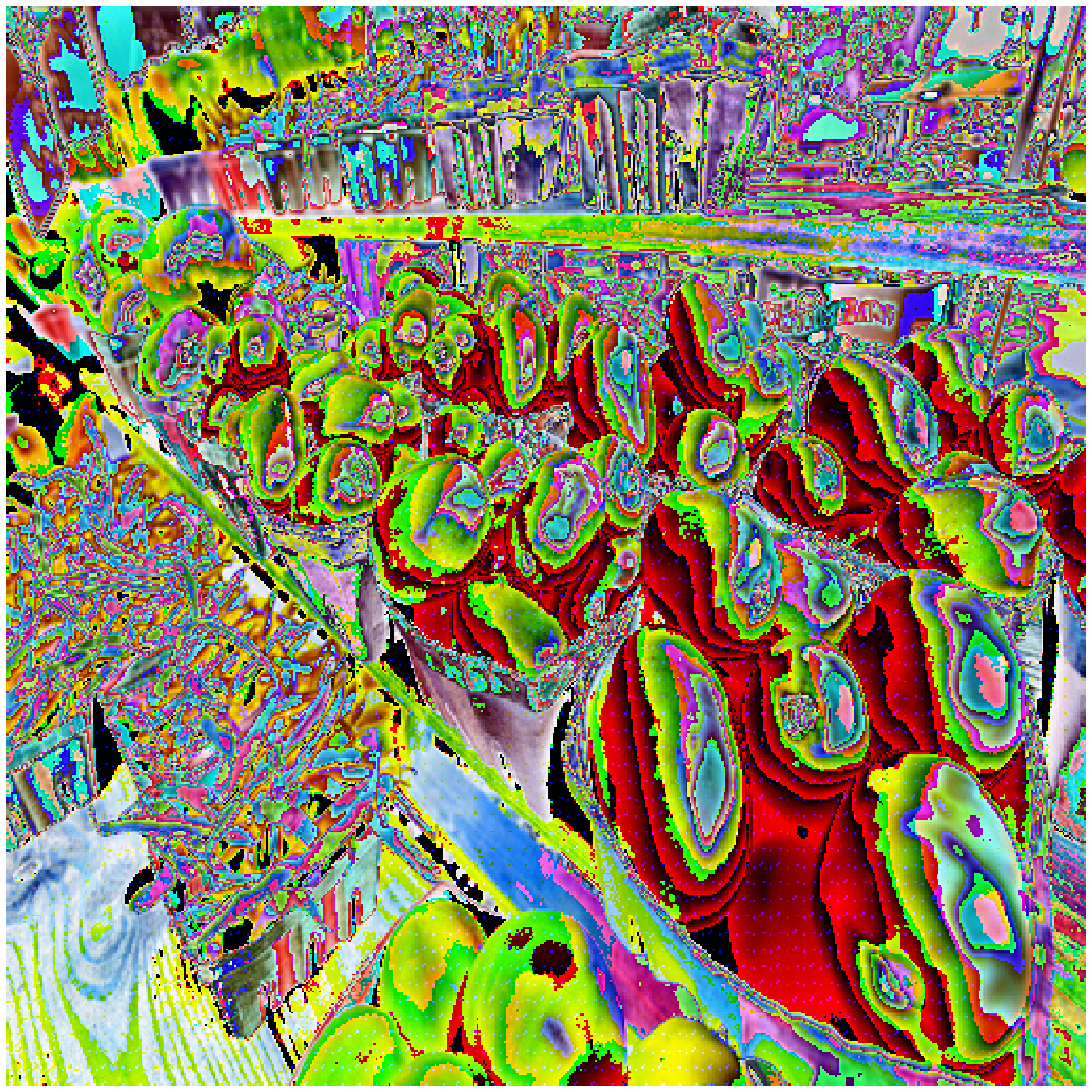}
		&
		\includegraphics[scale=3.62]{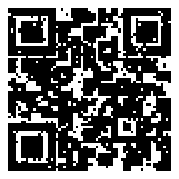}
		\\[.2cm]
		\includegraphics[scale=3.62]{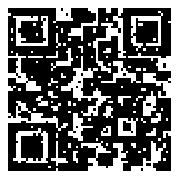}
		&
		\includegraphics[scale=3.54]{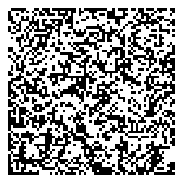}
	\end{tabular}
	\caption{Extracted watermarks under Gaussian Laplace attacks applied to NB0074.bmp watermarked image. The first row corresponds to CS. The second row corresponds to MS and KS, respectively.}
	\label{Gaussian_Laplace_noises}	
\end{figure}

\begin{figure}[H] % [ht]
	\centering
	\begin{tabular}{cc}
		\includegraphics[scale=.44]{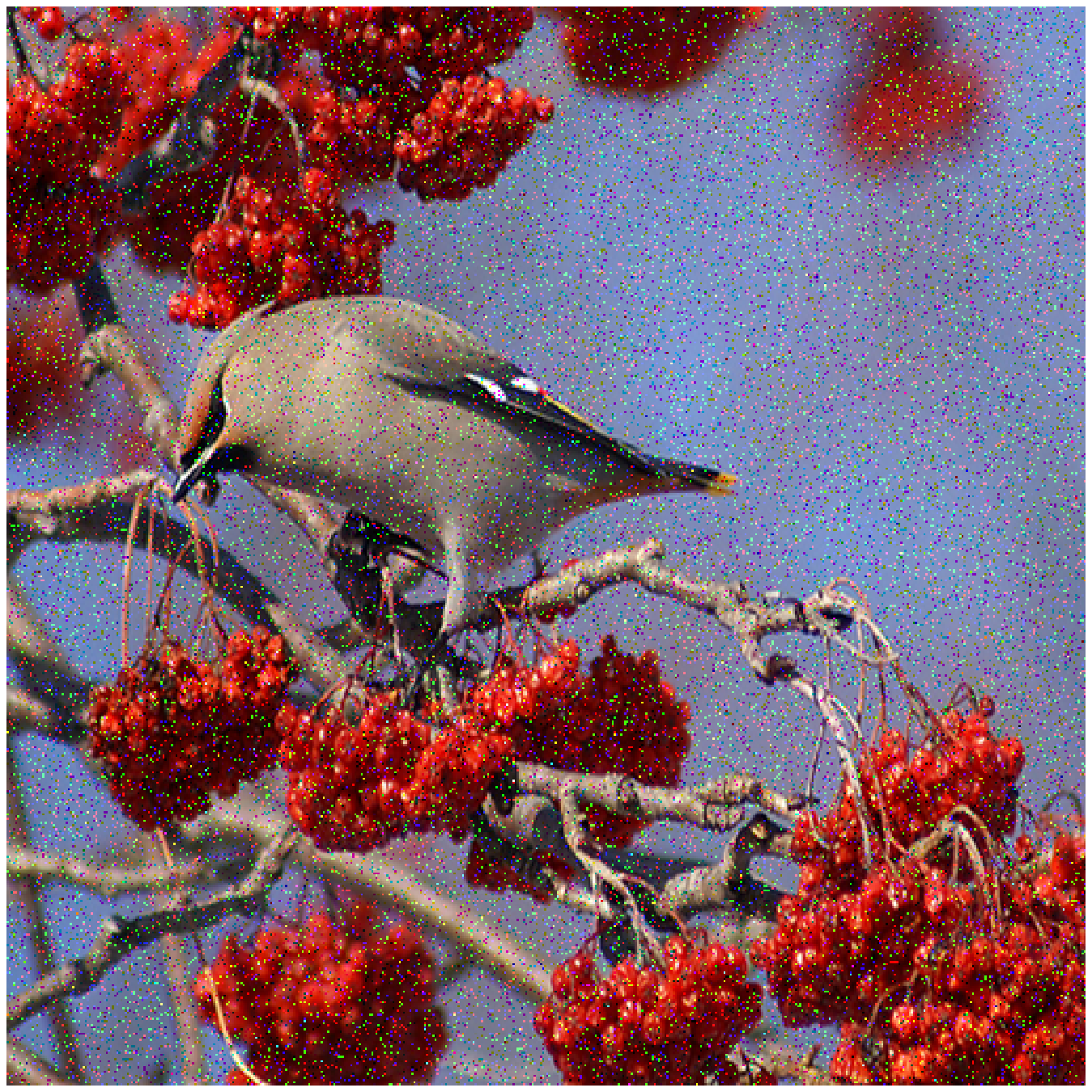}
		&
		\includegraphics[scale=3.62]{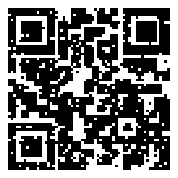}
		\\[.2cm]
		\includegraphics[scale=3.62]{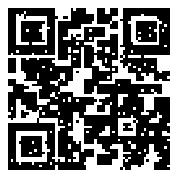}
		&
		\includegraphics[scale=3.62]{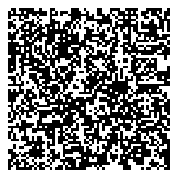}
	\end{tabular}
	\caption{Extracted watermarks under Salt and Pepper attacks applied to C0411.bmp watermarked image. The first row corresponds to CS. The second row corresponds to MS and KS, respectively.}
	\label{salt_pepper_noise}	
\end{figure}

The results of the experiment show that robustness of the watermarking scheme based on Charlier and Meixner Sobolev-type orthogonal polynomials is more remarcable than that obtained via Krawtchouk-Sobolev polynomials (see Figure \ref{Gaussian_Laplace_noises} and Figure \ref{salt_pepper_noise}). On the other hand, the graphs in Figure \ref{attacks} show that CS and MS methods remain be more robust against some of the noises, whereas KS seems to be more robust against Cropping and Gaussian attacks. These methods show to be similar in the sense of robustness when considering cropping attack.

\begin{figure}[H] % [ht]
	\centering
	\begin{tabular}{cc}
		\includegraphics[scale=.44]{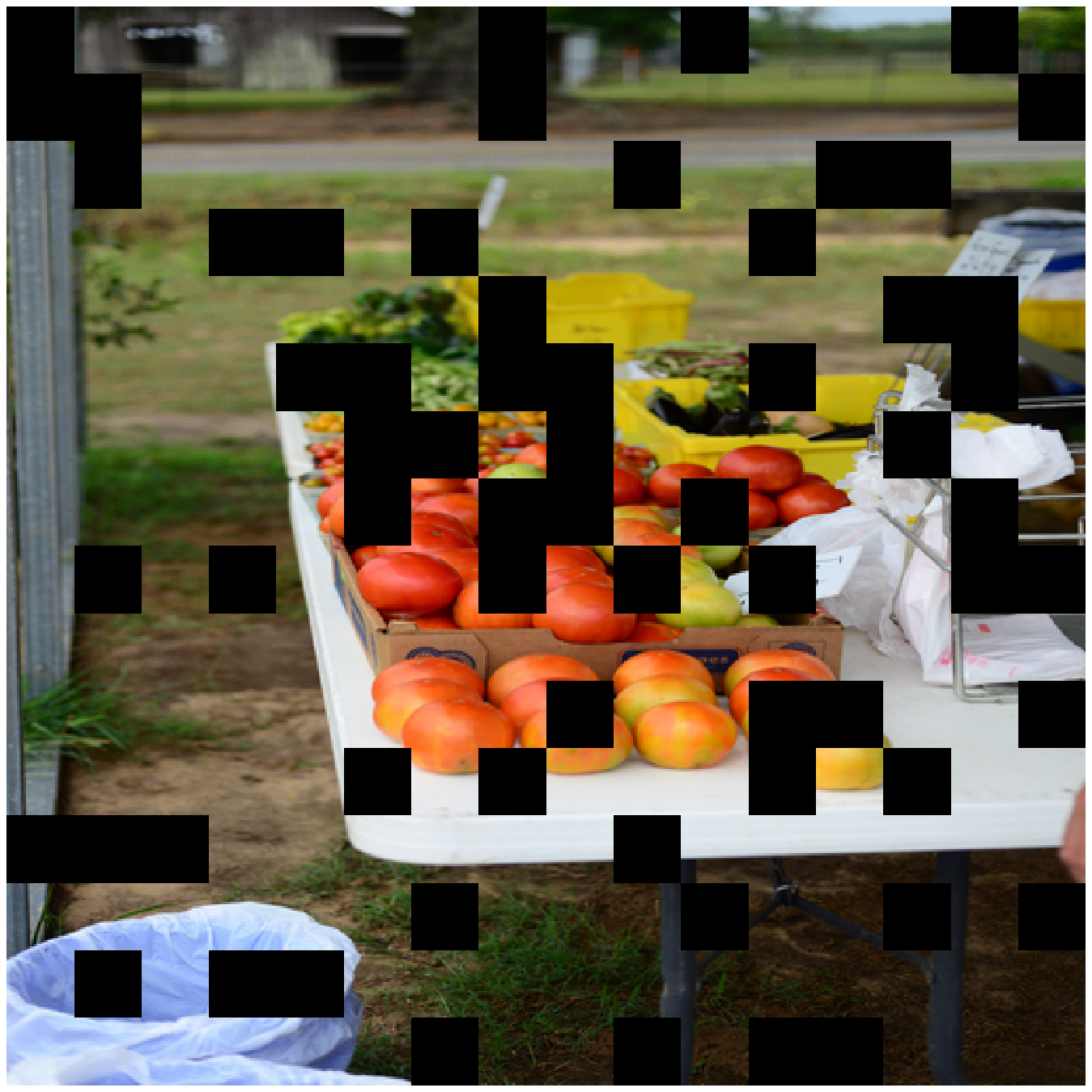}
		&
		\includegraphics[scale=3.62]{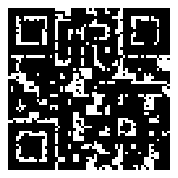}
		\\[.2cm]
		\includegraphics[scale=3.62]{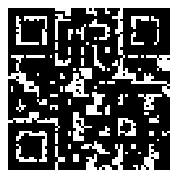}
		&
		\includegraphics[scale=3.62]{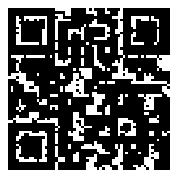}
	\end{tabular}
	\caption{Extracted watermarks under Cropping attacks applied to NB0206.bmp watermarked image. The first row corresponds to CS. The second row corresponds to MS and KS, respectively.}
	\label{Cropping_noise}	
\end{figure}

\subsection{Image tamper detection}
\begin{figure}[H] % [ht]
	\centering
	\begin{tabular}{cc}
		\includegraphics[scale=.44]{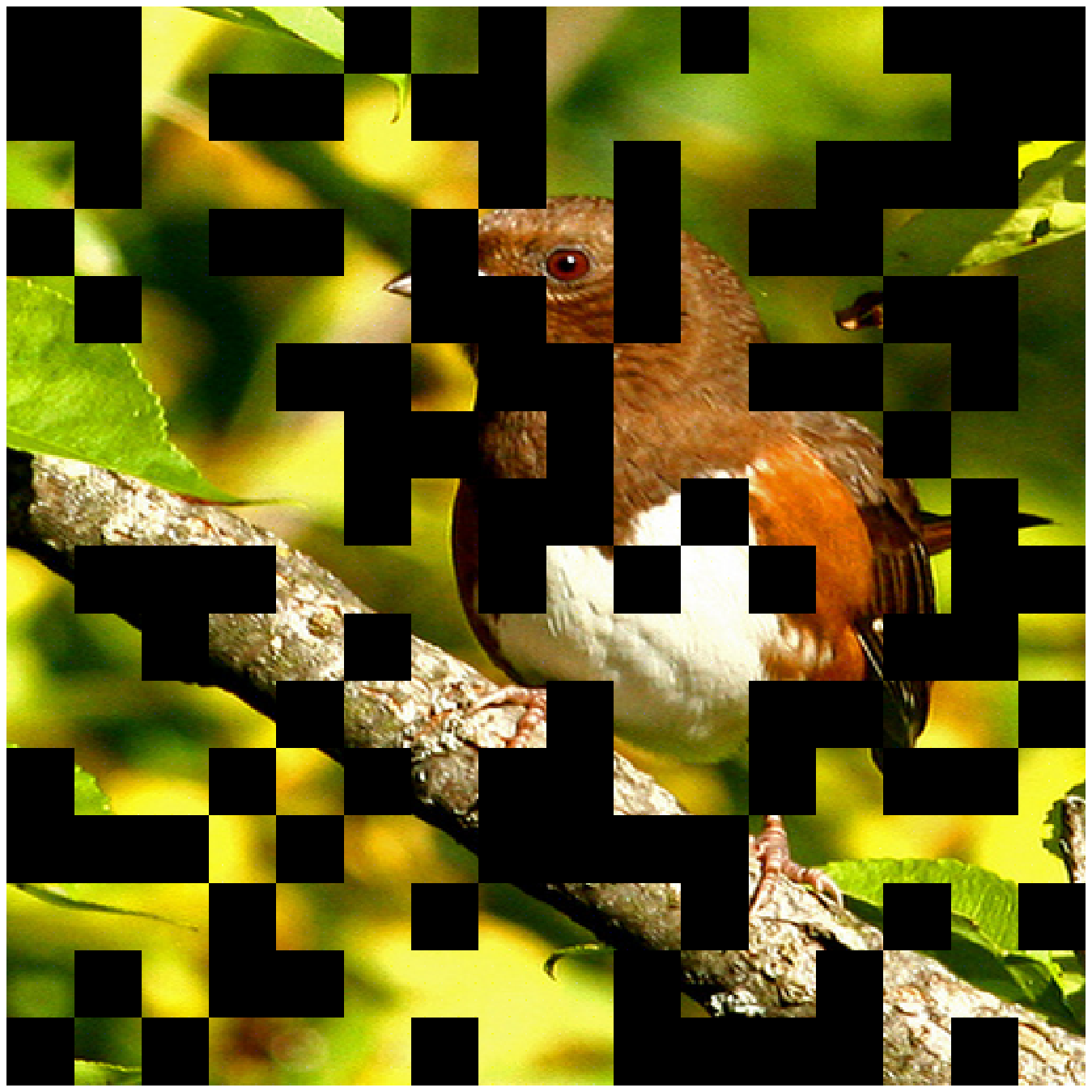}
		&
		\includegraphics[scale=.44]{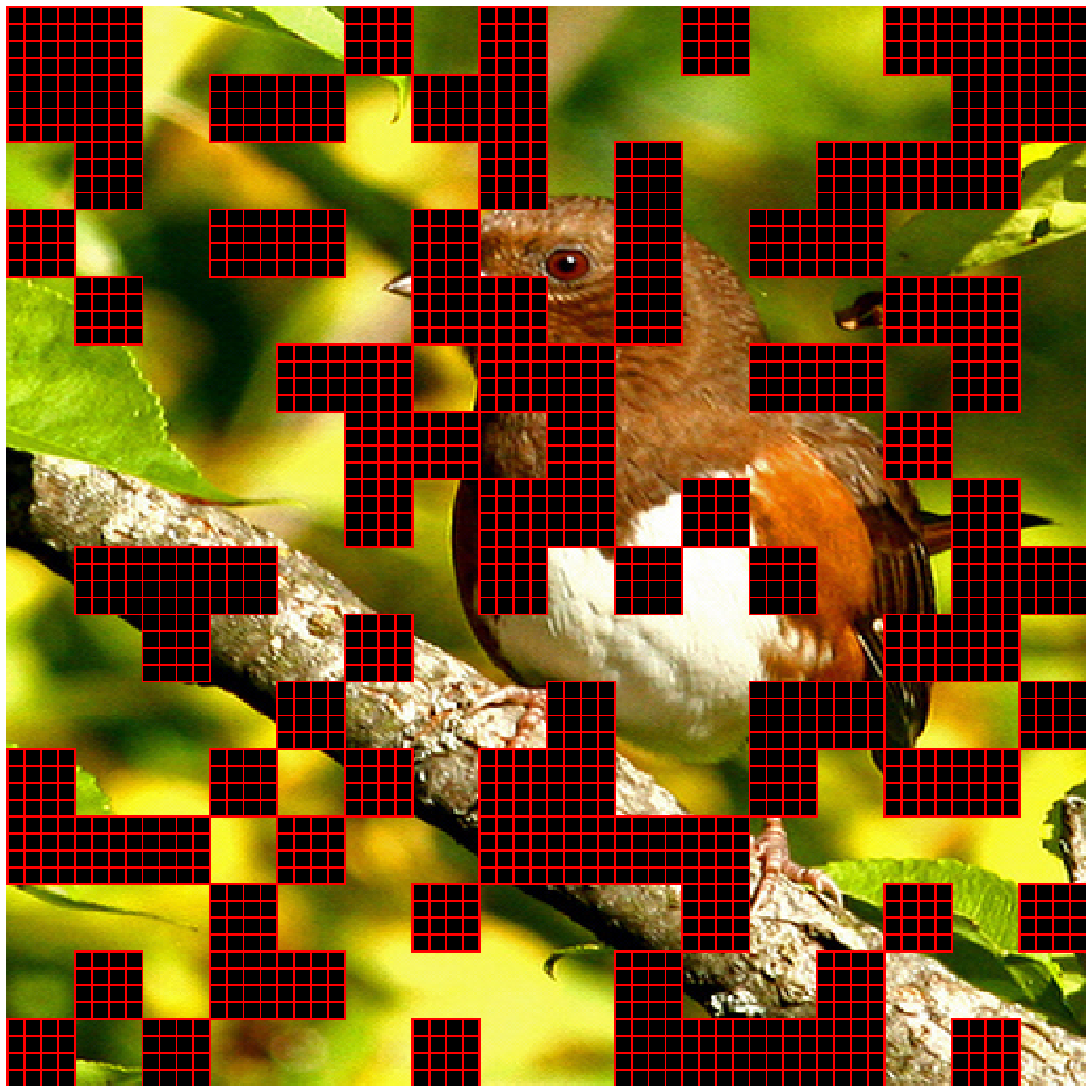}
		\\[.2cm]
		\includegraphics[scale=.44]{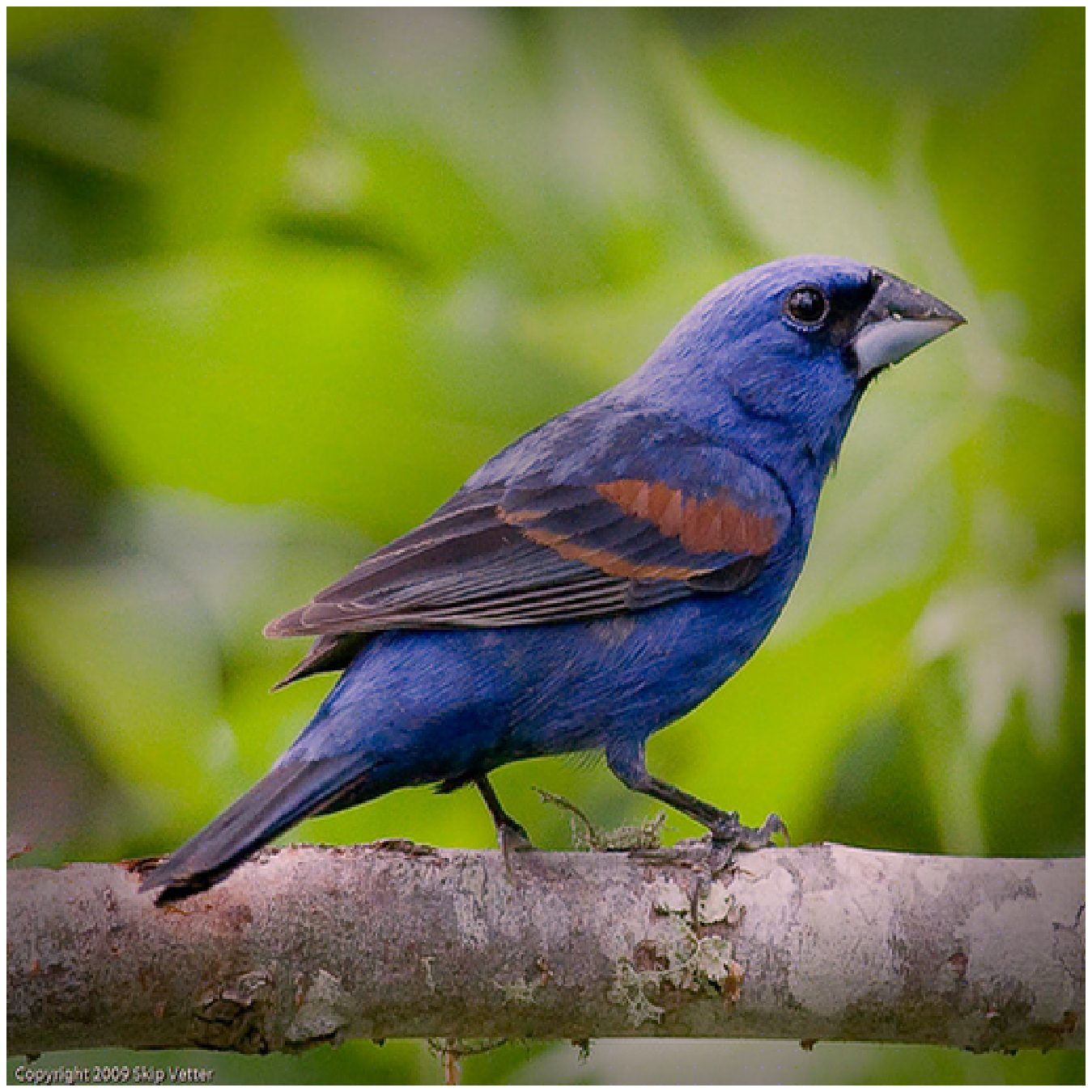}
		&
		\includegraphics[scale=.44]{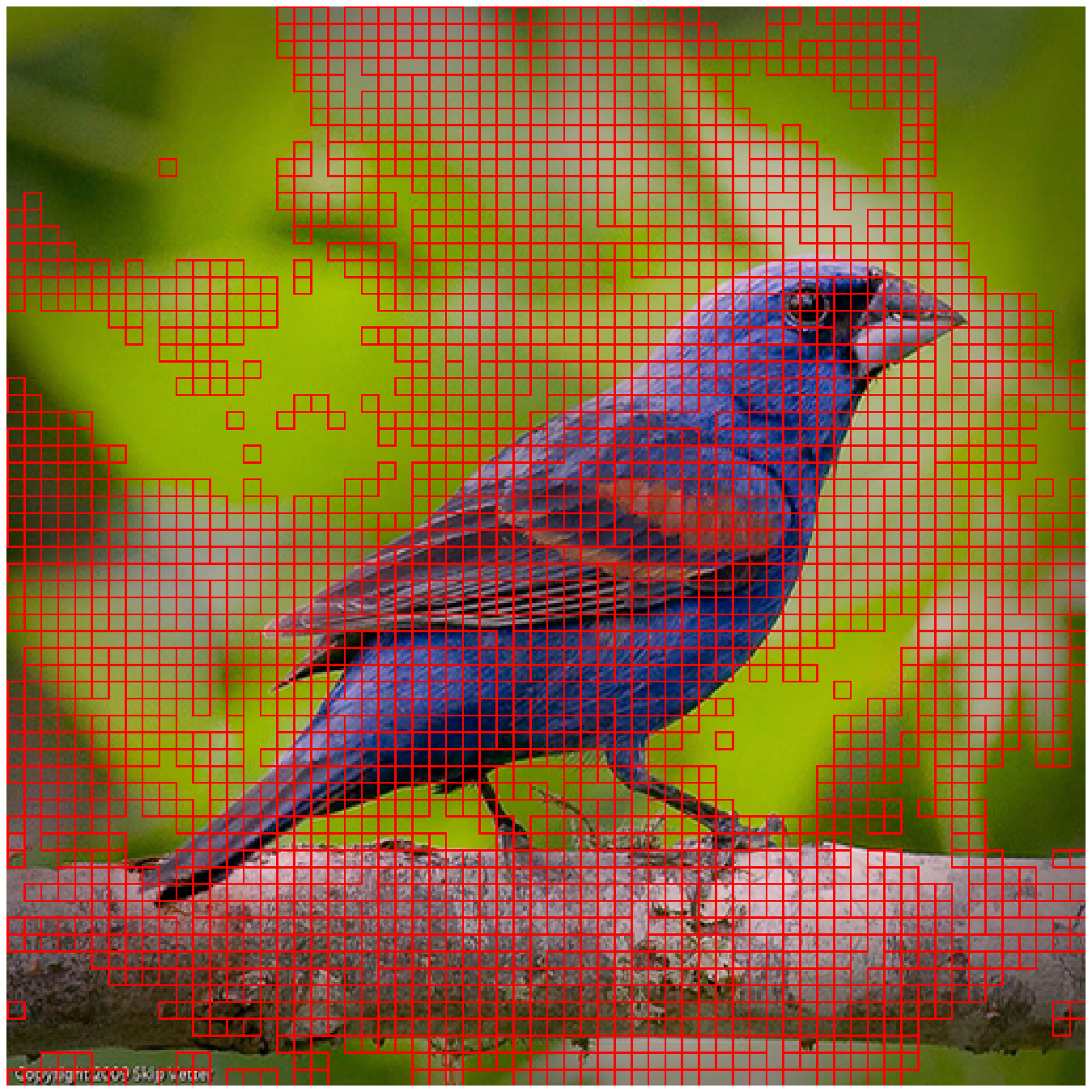}
	\end{tabular}
	\caption{Detection of the tampered regions of the C0886.bmp and C0309.bmp images corresponding to the Cropping and Fourier Ellipsoid noise}
	\label{tamperD}	
\end{figure}

Tampering is an intentional modification of documents in a way that would make them harmful for end users. It is essential to disclose through the fragile watermark any modifications made to the documents by unauthorized external agents. The modifications made were detected in one of the images considered from the tamper detection (see Figure \ref{tamperD}).

%%%%%%%%%%%%%%%%%%%%%%%%%%%%%%%%%%%%%%%%%%%%%%%%%%%%%%%%%%%%%%%%%%%%%%%%%%%%%%%%%%%%%%%%%%%%%%%%%%%%%%%

\section{Conclusions and further remarks}
In this research, a dual watermaking scheme based on Sobolev type orthogonal moments, has been proposed for the document authentication and the tamper detection. The performance of the method has been tested using six types of attacks to different regions of the documents with watermark. It was shown that robustness of the watermarking scheme based on CS and MS is much higher than that of the KS, proved to be of interest in watermarking procedures. In the tamper detection phase, the tampered regions of the watermarked document were detected. Experimental results illustrate that the proposed method produced watermarked document with a high visual quality and good PSNR values, which is in correspondence with the heuristic values of PSNR.

%%%%%%%%%%%%%%%%%%%%%%%%%%%%%%%%%%%%%%%%%%%%%%%%%%%%%%%%%%%%%%%%%%%%%%%%%%%%%%%%%%%%%%%%%%%%%%%%%%%%%%%

%%%%%%%%%%%%%%%%%%%%%%%%%%%%%%%%%%%%%%%%%%%%%%%%%%%%%%%%%%%%%%%%%%%%%%%%%%%%%%%%%%%%%%%%%%%%%%%%%%%%%%%
%%%%%%%%%%%%%%%%%%%%%%%%%%%%%%%%%%%%%%%%%%%%%%%%%%%%%%%%%%%%%%%%%%%%%%%%%%%%%%%%%%%%%%%%%%%%%%%%%%%%%%%

%%%%%%%%%%%%%%%%%%%%%%%%%%%%%%%%%%%%%%%%%%%%%%%%%%%%%%%%%%%%%%%%%%%%%%%%%%%%%%%%%%%%%%%%%%%%%%%%%%%%%%%

\section*{Declarations}

%%%%%%%%%%%%%%%%%%%%%%%%%%%%%%%%%%%%%%%%%%%%%%%%%%%%%%%%%%%%%%%%%%%%%%%%%%%%%%%%%%%%%%%%%%%%%%%%%%%%%%%

\begin{itemize}

\item Availability of data and material: There is no other data or extra material associated to this publication, but the presented in the manuscript.  

\item Conflict of interest: The authors declare no conflict of interest.

\item Funding: The work of A. Lastra and A. Soria-Lorente is partially supported by Dirección General de Investigación e Innovación, Consejería de Educación e Investigación of the Comunidad de Madrid (Spain) and Universidad de Alcalá, under grant CM/JIN/2021-014, Proyectos de I+D para Jóvenes Investigadores de la Universidad de Alcalá 2021. The work of A. Lastra is partially supported by the project PID2019-105621GB-I00 of Ministerio de Ciencia e Innovación, Spain; and by Ministerio de Ciencia e Innovación-Agencia Estatal de Investigación MCIN/AEI/10.13039/501100011033 and the European Union ``NextGenerationEU''/ PRTR, under grant TED2021-129813A-I00.

\item Contributions: All authors contributed equally to this work.

\item Acknowledgements: The work of A. Lastra and A. Soria-Lorente is partially supported by Dirección General de Investigación e Innovación, Consejería de Educación e Investigación of the Comunidad de Madrid (Spain) and Universidad de Alcalá, under grant CM/JIN/2021-014, Proyectos de I+D para Jóvenes Investigadores de la Universidad de Alcalá 2021. The work of A. Lastra is partially supported by the project PID2019-105621GB-I00 of Ministerio de Ciencia e Innovación, Spain; and by Ministerio de Ciencia e Innovación-Agencia Estatal de Investigación MCIN/AEI/10.13039/501100011033 and the European Union ``NextGenerationEU''/ PRTR, under grant TED2021-129813A-I00.

\end{itemize}

%%%%%%%%%%%%%%%%%%%%%%%%%%%%%%%%%%%%%%%%%%%%%%%%%%%%%%%%%%%%%%%%%%%%%%%%%%%%%%%%%%%%%%%%%%%%%%%%%%%%%%%

%%%%%%%%%%%%%%%%%%%%%%%%%%%%%%%%%%%%%%%%%%%%%%%%%%%%%%%%%%%%%%%%%%%%%%%%%%%%%%%%%%%%%%%%%%%%%%%%%%%%%%%

\end{document}